\documentclass[prd,superscriptaddress]{revtex4}
\usepackage{mathrsfs}
\usepackage{amsmath}
\usepackage{amssymb}
\usepackage{color}
\usepackage{graphicx}

\begin{document}

\title{Interpretation of the cosmic ray positron and antiproton fluxes}

\author{Paolo Lipari}
\email{paolo.lipari@roma1.infn.it}
\affiliation{INFN sezione Roma ``Sapienza''} 


\begin{abstract}
 The spectral shape of cosmic ray positrons and
 antiprotons has been accurately measured in the broad 
 kinetic energy range 1--350~GeV. In the higher part
 of this range ($E \gtrsim 30$~GeV) the $e^+$ and $\overline{p}$
 are both well described by power laws with spectral indices
 $\gamma_{e^+} \simeq 2.77 \pm 0.02$
 and $\gamma_{\overline{p}} \simeq 2.78\pm 0.04$
 that are approximately equal to each other and to the spectral
 index of protons. In the same energy range the
 positron/antiproton flux ratio has the approximately constant value
 $2.04 \pm 0.04$, that is consistent with being equal to the ratio
 $e^+/\overline{p}$ calculated for the conventional mechanism
 of production, where the antiparticles are created
 as secondaries in the inelastic
 interactions of primary cosmic rays with interstellar gas.
 The positron/antiproton ratio at lower energy is significantly
 higher (reaching a value $e^+/\overline{p} \approx 100$ for $E\approx 1$~GeV),
 but in the entire energy range 1--350~GeV,
 the flux ratio is consistent with being equal to ratio
 of the production rates in the conventional mechanism,
 as the production of low energy
 antiprotons is kinematically suppressed in collisions
 with a target at rest.
 These results strongly suggest that cosmic ray positrons and antiprotons
 have a common origin as secondaries in hadronic interactions.
 This conclusion has broad implications for the astrophysics of
 cosmic rays in the Galaxy.
\end{abstract}

\maketitle

 \clearpage

\section{Introduction}
\label{sec:introduction}

The study of cosmic ray positrons and antiprotons 
is a very important topic in high energy astrophysics and 
has received a very large amount of attention in recent years.
A strong motivation for these studies is the possibility
that, if the Galactic Dark Matter (DM) is in the form of
Weakly Interacting Massive Particles (WIMP's), the self--annihilation
or decay of the DM particles can be an observable source of relativistic
antiparticles.

The only known mechanism that can 
generate a detectable flux of high energy positrons and antiprotons
in the Galaxy is their production as secondaries in the inelastic interactions of
primary cosmic rays (CR) in interstellar space, or perhaps inside or around the
astrophysical sites where CR are accelerated. This conventional
mechanism has to be well understood to establish the existence
of other contributions generated by alternative
mechanisms (such as DM annihilation, or direct acceleration in some classes
of astrophysical sources).

The study of the CR fluxes of antiparticles has made a very important
step forward thanks to the measurements performed by 
PAMELA of positrons \cite{Adriani:2008zr,Adriani:2013uda}
and anti-protons \cite{pamela-antiprotons}.
The PAMELA observation showed that the positron flux, for $E \gtrsim 10$~GeV,
is significantly harder than predictions based on the
conventional mechanism, and this discrepancy
has generated an intense attention and a large body of literature.

The PAMELA results for the positron flux have been confirmed
by observations of the FERMI detector \cite{FermiLAT:2011ab}
performed using the Earth magnetic field to separate 
the fluxes of CR electrons and positrons.
More recently the AMS02 detector, on the International Space Station
has measured the $e^+$ spectrum \cite{Aguilar:2014mma}, in good agreement with
PAMELA, and with smaller errors and extending the observations
up to an energy $E \simeq 400$~GeV.

Most of the literature that discusses the CR positron flux
takes the point of view that the observations require a new,
non--standard source of positrons.
Only few works have taken the point of view that
the discrepancy between predictions and observations
must be attributed to an incorrect estimate of the positron flux generated by the
conventional (secondary production) mechanism.

In contrast to the situation for positrons,
the measurements of antiprotons performed by PAMELA are
consistent with the predictions based on the conventional mechanism of production.
This is clearly a very important constraint for the interpretation
of the antiparticle data,
because many sources of $e^+$ are also sources of $\overline{p}$
(and vice--versa). The comparison of positron and
antiproton fluxes is therefore crucial to determine the nature and properties of the
physical mechanisms that generate antiparticles in the Galaxy.

Recently, the AMS02 collaboration has published a new measurement
of  the cosmic  ray antiproton spectrum.
These results are in very good agreement
with the PAMELA measurements, but have smaller errors and extend to higher energy
(up to $E \simeq 350$~GeV). This allows to determine the shape
of the antiproton spectrum with much greater precision.

Having in hands precision measurements of the positron and antiproton
fluxes that extend in a broad energy range, together with high quality measurements
of the spectra of primary CR particles (protons, electrons and nuclei)
allows to perform detailed comparisons that can give important insight
about the origin of the different CR components.
This work is dedicated to a critical discussion of some
intriguing results that emerge from these comparisons.

For energies larger than approximately 30~GeV
the spectra of protons, electrons, antiprotons and positrons
are well described by power laws ($\phi_j (E) \propto E^{-\gamma_j}$).
It is remarkable that the spectral indices for positrons and antiprotons
are consistent with being equal
($\gamma_{e^+} \simeq 2.77 \pm 0.02$
and $\gamma_{\overline{p}} \simeq 2.78\pm 0.04$),
and are also very close to the spectral index for protons.
Only the electron flux has a different shape, and is much softer
with a larger spectral index of order $\gamma_{e^-} \approx 3.2$.

The $e^+/\overline{p}$ ratio is approximately constant 
in the energy range [30--350]~GeV with value $2.04 \pm 0.04$.
This value is also equal (taking into account systematic uncertainties)
to the $e^+/\overline{p}$ ratio {\em at production}
for the conventional mechanism.
In fact it is possible to show the numerical validity of
a stronger result: the observed $e^+/\overline{p}$ ratio,
is consistent with being equal to the ratio $e^+/\overline{p}$ calculated
for the conventional mechanism of secondary production, in the entire
energy range [1--350]~GeV, including the lower energy part where the
positron/antiproton ratio changes rapidly with $E$, taking the value
$e^+/\overline{p} \approx 100$ for kinetic energy $E \approx 1$~GeV.

These intriguing results, are perhaps just a numerical accident 
but, at least at face value, seem to indicate
that the CR fluxes of both positrons and antiprotons have their origin in
the conventional secondary production mechanism. This is in contrast
to all models where $e^+$ and $\overline{p}$ have different origin.
This interpretation has however some 
non trivial difficulties, and requires a deep revision of some important 
concepts for the formation and propagation of the Galactic cosmic rays.
In the following we will attempt to discuss critically these problems.

This work is organized as follows.
The next section summarizes recent measurements of the cosmic ray fluxes.
Section~\ref{sec:hadronic} discusses the production of positrons and antiprotons
in hadronic interactions, and computes the antiparticles production
rates in the solar neighborhood.
Section~\ref{sec:losses} discusess the energy losses of relativistic particles
propagating in the Galaxy.
Section~\ref{sec:formation} outlines some general concepts about the formation
of the CR fluxes in the Galaxy, 
and describes what can be considered the
current (broadly, but not universally accepted) ``standard framework''
for the study of the Galactic Cosmic Rays.
Section~\ref{sec:alternative} discusses alternative frameworks
for CR in the Galaxy that can explain naturally the correlations between
the positron and antiproton fluxes.
Section~\ref{sec:secondary-nuclei} contains a short discussion
of the significance of the measurements of secondary nuclei.
The final section gives a summary and comments of the most promising
directions for future studies to clarify the situation.
The appendix~\ref{sec:analytic} discusses an analytic estimate
for the ratio $e^+/\overline{p}$ at production under the assumption
that the antiparticles are generated as secondaries
in inelastic hadronic interactions.

\section{Cosmic Ray fluxes} 
\label{sec:fluxes}

\begin{figure} [hbt]
\begin{center}
\includegraphics[width=14.0cm]{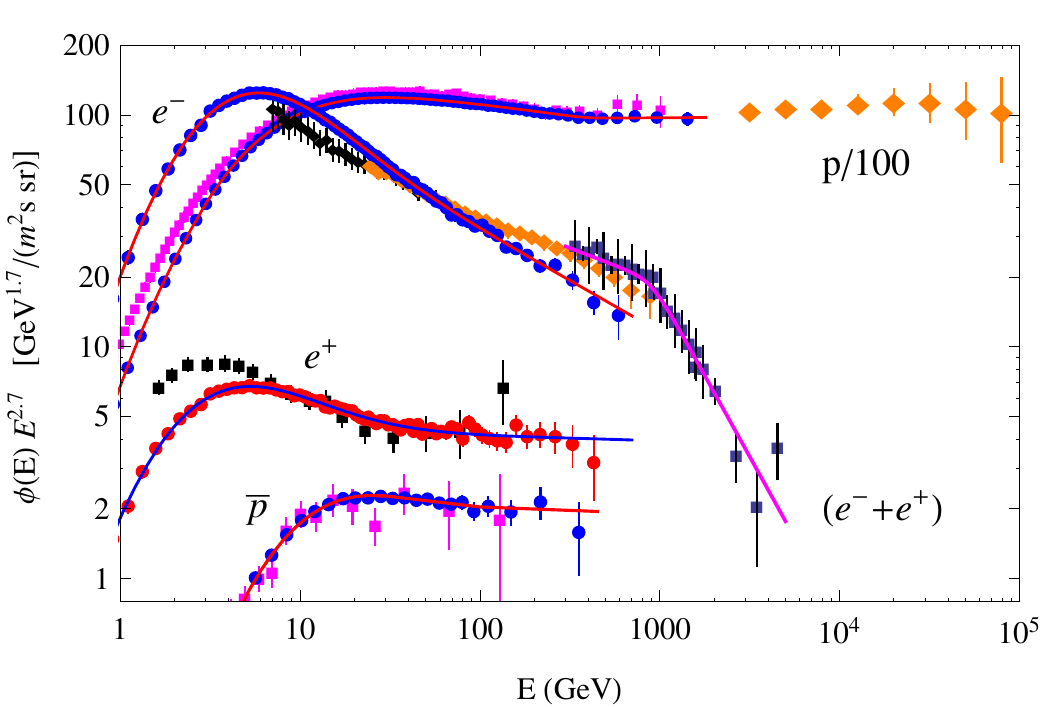}
\end{center}
\caption {\footnotesize
 Fluxes of $p$, $e^-$, $e^+$ and $\overline{p}$. The fluxes are shown in the form
 $\phi_j (E) \; E^{2.7}$ versus $E$ (with $E$ the kinetic energy)
 to enhance the features of the energy dependence.
 The proton data (rescaled by a factor $10^{-2}$)
 is from PAMELA (squares) \protect\cite{pamela-protons-helium},
 AMS02 (circles) \protect\cite{ams-protons} and
 CREAM (diamonds) \protect\cite{Yoon:2011aa}.
 The electron data is from AMS02 \protect\cite{Aguilar:2014mma}.
 The positron data is from PAMELA (squares)
 \protect\cite{Adriani:2013uda} and
 AMS02 (circles) \protect\cite{Aguilar:2014mma}.
 The ($e^- + e^+)$ data is from FERMI (diamonds) \cite{Abdo:2009zk} and HESS (squares)
\cite{Aharonian:2008aa,Aharonian:2009ah}.
The antiproton data is from PAMELA (squares) \protect\cite{pamela-antiprotons}
and AM02 (circles)\protect\cite{Aguilar:2016kjl}.
The lines are fit to the AMS02 data, and to the HESS $(e^- + e^+)$ data
\label{fig:allflux} }
\end{figure}

\begin{figure} [hbt]
\begin{center}
\includegraphics[width=14.0cm]{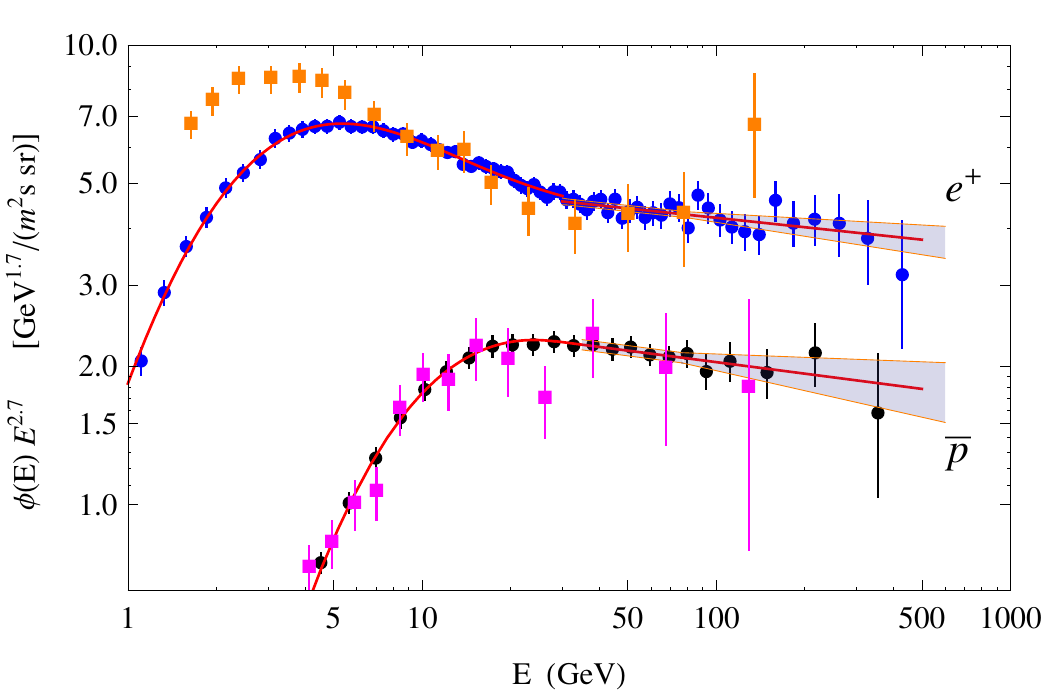}
\end{center}
\caption {\footnotesize
 Fluxes of $e^+$ and $\overline{p}$ shown in the form
 $\phi_j (E) \; E^{2.7}$ versus $E$.
 The circles are measurements by AMS02 \protect\cite{Aguilar:2014mma},
 \protect\cite{Aguilar:2016kjl}. 
 The squares are measurements by PAMELA \protect\cite{Adriani:2013uda} and
 \protect\cite{pamela-antiprotons}.
 The lines are smooth fits to the AMS02 data.
 The band indicate the range of power law fits to the AMS02 data
 in the range $E > 30$~GeV (see text).
\label{fig:antia-fit} }
\end{figure}

Recent measurements of the fluxes of protons, electrons, antiprotons and positrons
are shown in Fig.~\ref{fig:allflux} in the form $\phi_j (E) \times E^{2.7}$ 
versus $E$ with $\phi_j(E)$ the differential energy spectrum for
particle type $j$ and $E$ the kinetic energy. The spectra are shown
multiplied by the factor $E^{2.7}$ for a better visualization of the
spectral shapes, note also that the proton flux is rescaled by a factor $10^{-2}$
The measurements are by PAMELA (for $p$, $\overline{p}$ and $e^\mp$)
\cite{pamela-protons-helium,pamela-antiprotons,Adriani:2013uda,Adriani:2011xv}, 
AMS02 (also $p$, $\overline{p}$ and $e^\mp$) \cite{ams-protons,Aguilar:2014mma,Aguilar:2016kjl}
CREAM ($p$ at high energy) \cite{Yoon:2011aa},
FERMI ($e^+ + e^-$)
\cite{Abdo:2009zk} and HESS ($e^+ + e^-$) 
\cite{Aharonian:2008aa,Aharonian:2009ah}.
 The measurements of the $e^+$ and $\overline{p}$ spectra are 
 also shown with an enlarged scale in Fig.~\ref{fig:antia-fit}.

 In this work the spectra of different particle types are
 shown and compared as differential distribution in kinetic energy. 
 This choice is not unique. An equally valid possibility is to
 show and compare magnetic rigidity spectra at the same (absolute) value
 of the rigidity.
 The following discussion about the mechanisms that shape the
 CR (anti)--particle spectra can be easily reformulated
 for a different choice of the variable used to represent
 the observations. Obviously, for particles of unit charge the energy and the
 (absolute value of the) rigidity become equal in the limit of large $E$ .

\begin{figure} [hbt]
\begin{center}
\includegraphics[width=14.0cm]{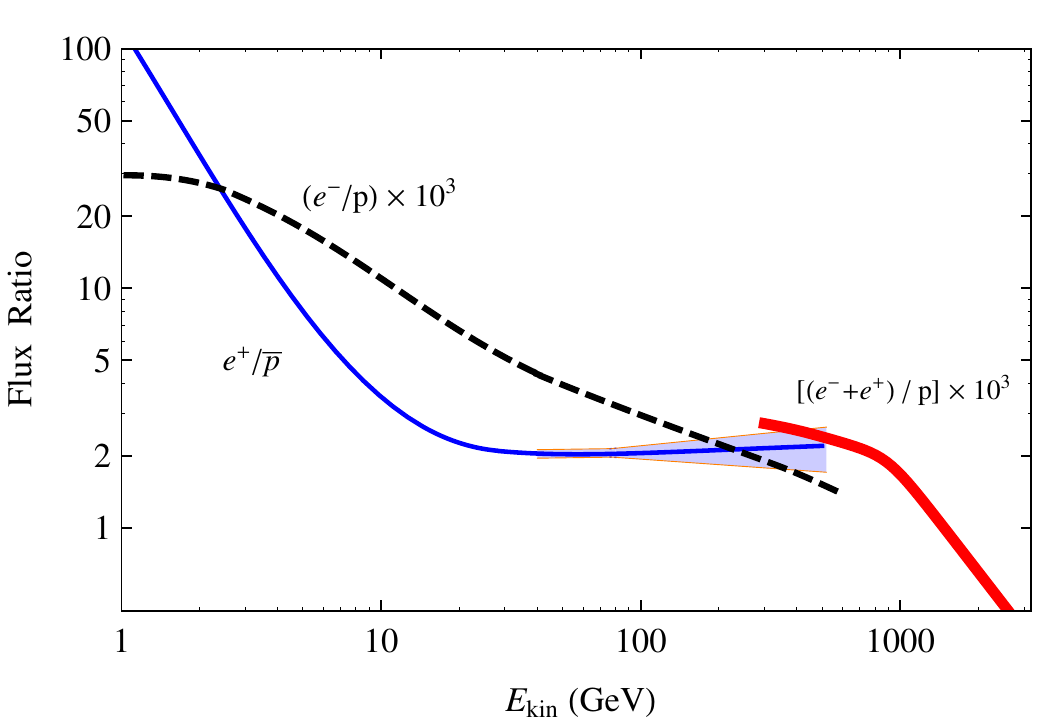}
\end{center}
\caption {\footnotesize
 Ratios $e^+/\overline{p}$, $e^-/p$, and $(e^-+e^+)/p$ between the fluxes
 of different particle types.
 All lines are ratios of the fits to the experimental data 
 shown in Fig.~\ref{fig:allflux} and Fig.~\ref{fig:antia-fit}.
 For the $e^+/\overline{p}$ ratio the fit given in equation (\ref{eq:phi-ratio}) is also
 shown.
 \label{fig:ratio-ep} }
\end{figure}

\begin{figure}[hbt]
\begin{center}
\includegraphics[width=14.0cm]{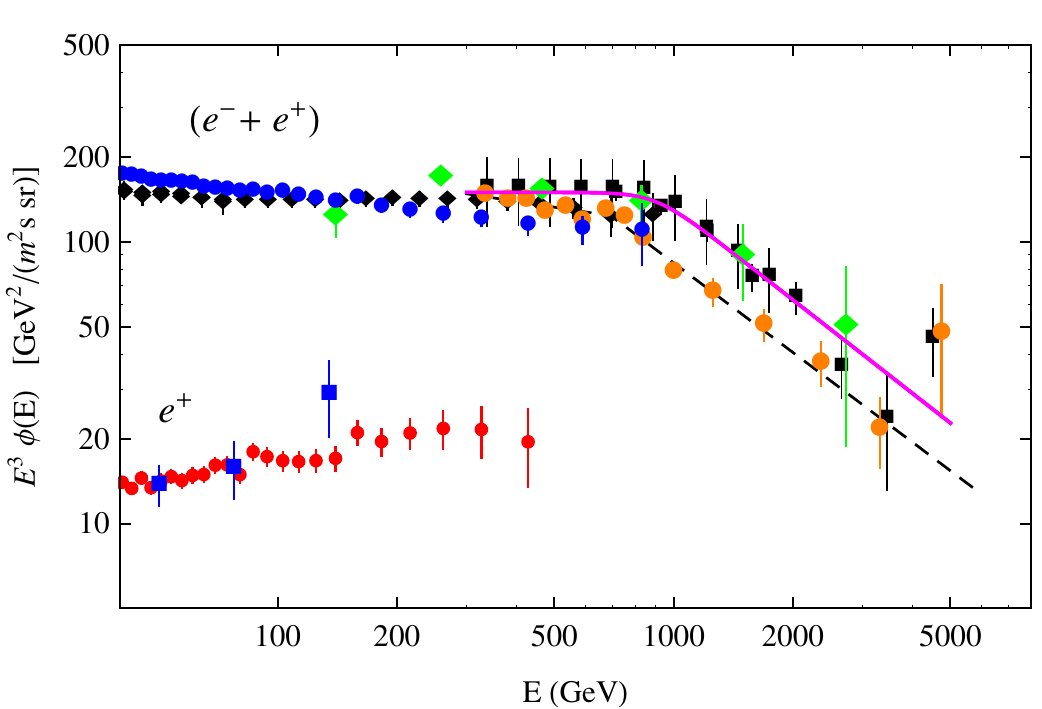}
\end{center}
\caption {\footnotesize
 Flux measurements of the sum ($e^+ + e^-)$ obtained by AMS02 (small circles)
 \protect\cite{Aguilar:2014fea}, FERMI (small diamonds),
 HESS (squares), MAGIC (diamonds) and VERITAS (big circles).
 The solid (dashed) line is the best fit to the data obtained by
 HESS \protect\cite{Aharonian:2009ah} (VERITAS \protect\cite{Staszak:2015kza}). 
\label{fig:lept2}}
\end{figure}

\vspace{0.2 cm}
Inspecting Fig.~\ref{fig:allflux} and~\ref{fig:antia-fit}
one can make the following remarks:

 \begin{itemize}

 \item The proton flux for $E \gtrsim 30$~GeV has approximately a power law form
 ($\phi_p (E) \propto E^{-\gamma_p}$)
 with an intriguing hardening feature at energy $E \simeq 300$~GeV.
 The feature was first deduced by CREAM \cite{cream-discrepant-hardening},
 then directly observed by PAMELA \cite{pamela-protons-helium}
 and now is also clearly confirmed by the AMS02 data \cite{ams-protons}.
 The AMS02 collaboration \cite{ams-protons} has fitted its measurement
 of the proton flux as the transition between two power laws
 with exponents
 $\gamma_{p}$ and $\gamma_{p} + \Delta \gamma_p$ above and below the
 transition momentum $p^* \simeq 336^{+44}_{-26}$~GeV. The best fit values
 of the parameters are: 
 $\gamma_{p} \simeq 2.849^{+0.006}_{-0.005}$
and
 $\Delta \gamma_p = -0.133^{+0.056}_{-0.037}$.

 \item 
 The energy spectrum of cosmic ray electrons, 
 is significantly smaller and softer than the 
 corresponding proton spectrum.
 The ratio $e^-/p$ of the electron and proton
 fluxes is shown in Fig.~\ref{fig:ratio-ep} as a function of the 
 particle kinetic energy $E$.
 For low energy ($0.5 \lesssim E \lesssim 2$~GeV)
 the ratio $e^-/p$ has a value of order 0.03.
 Increasing $E$ the ratio falls rapidy, and 
 at $E= 100$~GeV it is approximately $3\times 10^{-3}$, one order
 of magnitude smaller.
 The energy dependence of the ratio in the energy range
 $E \in [30,300]$~GeV is reasonably well described by a power law:
 \begin{equation}
 \left . \frac{\phi_{e^-} (E)}{\phi_{p}(E) } \right |_{E \in [30,300]~{\rm GeV}}
 \simeq
 (3.95 \pm 0.10) \times 10^{-3} \;
 \left (
 \frac{E}{50~{\rm GeV}} \right )^{-0.41\pm 0.02}
 \label{eq:phi-ratioele}
 \end{equation}

 \item Inspecting Fig.~\ref{fig:allflux} and~\ref{fig:antia-fit} 
 it is apparent that for kinetic energy $E \gtrsim 20$--30~GeV
 the spectra of positrons and antiprotons
 have a shape that is reasonably well approximated by a simple power law.
 To test this hypothesis, we have fitted the AMS02 data points
 in the energy range $E > 30$~GeV with the functional form:
 \begin{equation}
 \phi_j(E) = K_j \; \left (\frac{E}{50~{\rm GeV}} \right )^{-\gamma_j}
 \end{equation}

 (with $j = e^+$ or $\overline{p}$).
 For positrons we obtain a best fit with $\chi^2_{\rm min} = 12.0$ for
 27 degrees of freedom (d.o.f.), 
 and best fit parameters
 $K_{e^+} = (11.4 \pm 0.1) \times 10^{-5}$~(m$^2$s\,sr\,GeV)$^{-1}$ and 
 $\gamma_{e^+} = 2.77 \pm 0.02$.
 The $\chi^2$ of the fit has been calculated summing in quadrature statistical and
 systematic errors, the resulting value of $\chi^2_{\rm min}$ is small,
 indicating the existence of
 correlations in the systematic errors for different points.
 For antiprotons the best fit has $\chi^2_{\rm min} = 1.56$
 for 10 d.o.f. with best fit parameters
 $K_{\overline{p}} = (5.6 \pm 0.1)\times 10^{-5}$~(m$^2$s\,sr\,GeV)$^{-1}$, and 
 $\gamma_{\overline{p}} = 2.78 \pm 0.04$.

 It is remarkable that the estimate of the (approximately constant)
 spectral indices of the positron and anti-proton spectra 
 in the energy range $E \in [30,350]$~GeV are consistent with being equal.
 It is also intriguing that the spectral indices of the antiparticle
 fluxes are very close to the spectral index for protons:
\begin{equation}
\gamma_{e^+} \simeq \gamma_{\overline{p}} \approx \gamma_p + \frac{\Delta \gamma_p}{2}
\label{eq:gamma1}
\end{equation}
In the following we will discuss if these results are a simple coincidence
or have a deeper physical explanation.

\item The positron/antiproton ratio
 can be estimated combining the results of the two fits:
 \begin{equation}
 \left . \frac{\phi_{eˆ+} (E)}{\phi_{\overline{p}}(E) } \right |_{E \in [30,350]~{\rm GeV}}
 \simeq
 (2.04 \pm 0.04) \times
 \left (
 \frac{E}{50~{\rm GeV}} \right )^{0.015\pm 0.045}
 \label{eq:phi-ratio}
 \end{equation}
 that is consistent with a constant value.
 This result is in striking contrast with the behavior of 
 the $e^-/p$ ratio (given in equation (\ref{eq:phi-ratioele})) that in the same
 energy range falls rapidly with energy. 

 The energy dependence of the $e^+/\overline{p}$ ratio
 in a broader energy range is shown in Fig.~\ref{fig:ratio-ep}.
 For low kinetic energy the ratio varies rapidly with $E$.
 At $E \simeq 1$~GeV it is of order 100,
 and decreases monotonically to reach the approximately
 constant value of order 2 at $E \simeq 20$--30~GeV.

\item
Measurements of the spectrum of the sum $(e^+ + e^-)$
at very high energy have been obtained by the 
ground based Atmospheric Cherenkov telescopes
HESS \cite{Aharonian:2008aa,Aharonian:2009ah},
MAGIC \cite{BorlaTridon:2011dk}
and VERITAS \cite{Staszak:2015kza}.
The results of three Cherenkov telescopes are shown in Fig.~\ref{fig:lept2}.

These results show the existence of a sharp softening
of the $(e^+ + e^-)$ spectrum at an energy just below 1~TeV.
The HESS Collaboration has published a best fit to the data using a broken
power with break at energy $E\simeq 900$~GeV,
where the spectral index changes from $\gamma_1 \simeq 3.0$ to $\gamma_2 \simeq 4.1$
(no errors are reported).
The VERITAS \cite{Staszak:2015kza}) has also performed a fit using
the same broken power law functional form,
finding the break at $E \simeq 710 \pm 40$~GeV
and spectral indices $\gamma_1 \simeq 3.2 \pm 0.1$ and $\gamma_2 \simeq 4.1 \pm 0.1$
(all errors are only statistical).
The origin of this very prominent feature remains controversial
(see discussion in following).
 \end{itemize}

\section {Hadronic Interactions and the production of $e^+$ and $\overline{p}$}
\label{sec:hadronic}
The ``normal'' mechanism for the production of relativistic positrons and antiprotons
is their creation as secondaries in the inelastic interactions of primary cosmic rays
with some target material. The leading contribution to this production mechanism
is due to $pp$ interaction where one relativistic proton collides with
a target proton at rest. Additional contributions are due to helium--proton,
proton--helium, helium--helium, $\ldots$ collisions, where different types of CR
primary particles interact with different target nuclei.
In inelastic hadronic collisions baryon/antibaryon pairs can be created, and after
chain decay each antibaryon generates one stable antiproton that enters the CR
population. The most abundant source of positrons in hadronic interactions
is the creation and chain decay of $\pi^+$:
 \begin{equation*}
 \pi^+ \to \nu_\mu + \mu^+ \to \nu_\mu + [e^+ + \nu_e + \overline{\nu}_\mu ] ~.
 \end{equation*}
The spectrum of the positrons generated in this decay can be calculated
exactly using the (non--trivial) matrix element for the 3--body muon decay, 
and taking into account the fact that the muons created in the first decay
are in a well defined polarization state (see for example \cite{Lipari:2007su}).
A subdominant, but not entirely negligible, source of positrons is the 
production and chain decay of kaons. Also in this case
it is straightforward to compute the positron spectrum generated
by the decay of a kaon, summing over all chain decay modes
(weighted by the appropriate branching ratios).
For example a $K^+$ can generate positrons
either directly (in the decay mode $K^+ \to e^+ \nu_e \pi^\circ$)
or after several chain decay modes (such as in $K^+ \to \mu^+ \nu_\mu\to e^+ \ldots$).

The leading ($pp$) contribution to the local production rate
(that is the number of particles created per unit time, unit volume and
unit energy) of positrons (or antiprotons) can be written explicitely as:
 \begin{equation}
 q_{pp\to e^+ (\overline{p})}^{\rm loc} (E) = n_{{\rm ism},p} (\vec{x}_\odot)
 ~\int_{E}^\infty dE_0 ~n_p^{\rm loc} (E_0) \; \beta \, c \, \sigma_{pp} (E_0)
 ~ \left [ \frac{dN}{dE} (E, E_0) \right ]_{pp \to e^+ (\overline{p})} ~.
 \label{eq:qpp}
 \end{equation}
 In this equation
$n_{{\rm ism},p} (\vec{x}_\odot)$ is the density of protons in
interstellar gas in the solar neighborhood,
$n_p^{\rm loc}(E_0)$ is the number density
of CR protons with energy $E_0$ in the vicinity of the Sun,
$\beta \, c$ is the proton velocity, 
$\sigma_{pp} (E_0)$ is the $pp$ inelastic cross section (for collisions with
a proton at rest), and $dN_{pp\to j}/dE$ is the inclusive spectrum of
secondary particles of type $j$ in the final state of the collision,
calculated allowing the chain decay of all unstable parents.
The density in interstellar space is so low that it is safe to assume
that all unstable particles decay without suffering energy losses.
The integration is over all possible energies of the interacting proton.

The other contributions to the production
of $e^+$ and $\overline{p}$ (such as $p$--helium or helium--$p$) can be
obtained with obvious substitutions.

In the following we will show the results of a complete calculation
of the local production rate of positrons and antiprotons, performed
integrating numerically equation (\ref{eq:qpp}) and the other contributions.
The calculation is straightforward and requires three basic elements:
\begin{itemize}
\item[(i)]
 An estimate of the density (and composition) of interstellar gas in
 the solar neighborhood. 
\item[(ii)]
 An estimate of the fluxes of CR protons and nuclei in the local interstellar medium.
\item[(iii)]
 A model for the properties of inelastic hadronic interactions. 
\end{itemize}

For the density of the interstellar gas we have used the value $n_{\rm ism} = 1$~cm$^{-3}$
and assumed a standard composition with a helium contribution of 9\%.

\begin{figure} [hbt]
\begin{center}
\includegraphics[width=14.0cm]{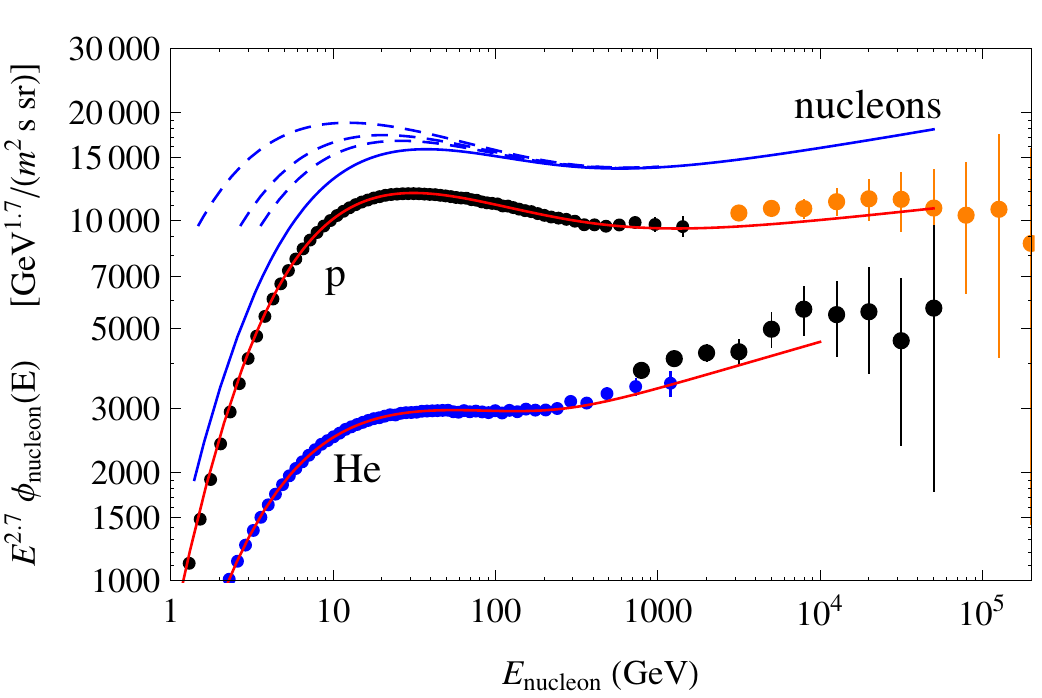}
\end{center}
\caption {\footnotesize
Interstellar spectra for protons and helium used in this work.
The data points are measurements performed by AMS02 (small circles)
\cite{ams-protons,Aguilar:2015ctt}
and CREAM (big circles) \cite{Yoon:2011aa}.
The lines show the all nucleon flux at the Earth) solid line
and in interstellar space. To correct for solar modulation effects
the potential has been estimated as $V = 0.4$, 0.6 and 0.9~GV.
\label{fig:primary} }
\end{figure}

The CR density in the local interstellar space can be easily related to the
flux $\phi_j^{\rm loc} (E)$ using the (very good) approximation of isotropy.
For example for protons:
\begin{equation}
n_p^{\rm loc} (E) = \frac{4 \, \pi}{\beta \, c} \; \phi_p^{\rm loc} (E) ~.
\end{equation}
A more difficult step is the estimate of the local interstellar spectra
from the CR measurements performed near the Earth deconvolving the effects of
the (time dependent) solar modulations. 
The model of the cosmic ray fluxes used in this calculation is shown in
figure~\ref{fig:primary} in the form of the all--nucleon flux versus the kinetic energy
per nucleon. The model is based on fits to the AMS02
proton and helium data \cite{ams-protons,Aguilar:2015ctt}
with a contribution of order 10\% from heavier nuclei
estimated from the data of the HEAO detector \cite{Engelmann:1990zz}.
At higher energy it joins smoothly with the model of Gaisser, Stanev and Tilav
\cite{Gaisser:2013bla}.

To unfold the effects of solar modulations, we have assumed that the modulations
is described by the Force Field Approximation (FFA) \cite{Gleeson:1968zza}.
In the FFA the solar modulations depend on one (time dependent) parameter, the
potential $V(t)$. The physical meaning of $V(t)$ is that particles with electric charge $Z$
that penetrate the heliosphere and reach the Earth lose the energy $\Delta E = |Z| \, V(t)$.
This is sufficient (using the Liouville theorem) to compute the observable flux
at the Earth, and the algorithm can also be inverted analytically \cite{Lipari:2014gfa}.
To take into account theoretical uncertainties we have
calculated the CR spectra in the local interstellar space deconvolving
the CR measurements shown in Fig.~\ref{fig:primary} with 
three values of the potential ($V = 0.4$, 0.6~and 0.9~GeV).
This gives a non negligible range of uncertainty for primary energies
$E \lesssim 50$~GeV. At higher energy the corrections associated to the
solar modulations become negligible.

In order to compute the spectra of secondary positrons and antiprotons
with an energy as large as $E \simeq 400$~GeV it is necessary to have a description
of the primary fluxes up to an energy per nucleon of order 40~TeV.
This requires an extrapolation of the measurements performed by the
magnetic spectrometers, and requires a modeling
of the hardening feature observed at a rigidity of order 250--300~GV. This
is an important source of uncertainty.

The final element required for the calculation of the
production rates of antiparticles is a model for the inclusive spectra
of secondaries in inelastic hadronic interactions. This is probably
the main source of systematic uncertainty in the calculation.
For a calculation that covers the entire energy range $E \gtrsim 1$~GeV,
we have used simple analytic descriptions of the differential cross 
sections for the production of secondary hadrons in $pp$ interactions
constructed on the basis of fixed target accelerator data.
For $\overline{p}$ production we have used the 
parametrization of Tan and Ng \cite{Tan:1982nc}
assuming also that the interactions generate equal spectra 
of antiprotons and antineutrons. 
For the production of charged pions we used the parametrizations of 
Badhwar et al. \cite{Badhwar:1977zf}, for kaons
the work of Anticic et al. \cite{Anticic:2010yg}.

For a cross check, and to have a first order estimate 
of the size of the uncertainties we have also performed a calculation, 
using the Monte Carlo code Pythia (version 6.4) \cite{Sjostrand:2006za}
to obtain numerically the spectra of secondaries.
The Pythia code is reliable only at sufficiently high energy, and we
have performed this calculation only for (positron and antiproton)
energy $E \ge 100$~GeV.

\begin{figure} [hbt]
\begin{center}
\includegraphics[width=14.0cm]{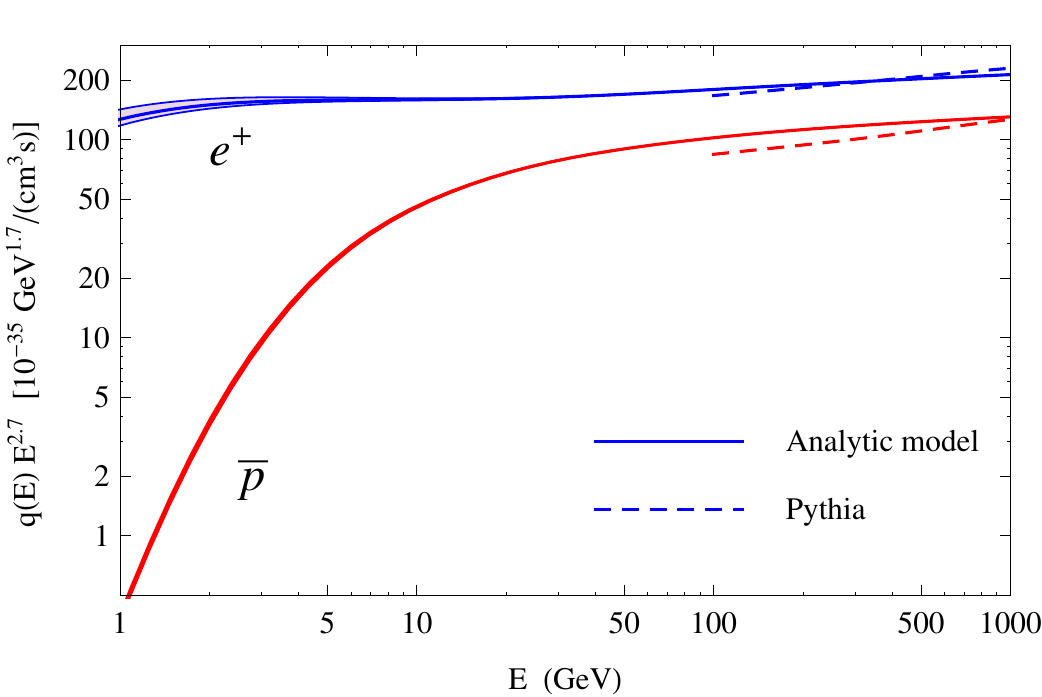}
\end{center}
\caption {\footnotesize
 Production rates of $\overline{p}$ and $e^+$ calculated with the 
 conventional mechanism in the solar neighborhood.
 The production rates have been calculated using
 two methods: an analytic description of the differential hadronic
 cross sections
 \protect\cite{Tan:1982nc,Badhwar:1977zf,Anticic:2010yg},
 and (for $E > 100$~GeV) the Pythia Monte Carlo code
 \protect\cite{Sjostrand:2006za}.
\label{fig:injection} }
\end{figure}

The results of the calculation of the local production rates of $e^+$ and $\overline{p}$
are shown in Fig.~\ref{fig:injection}.
plotted in the form $q_j (E) \times E^{2.7}$ versus the kinetic energy $E$
Some interesting features of the numerical results are listed below
(in the following of this section we will drop the superscript ``loc'' in the notation).

\begin{figure} [hbt]
\begin{center}
\includegraphics[width=14.0cm]{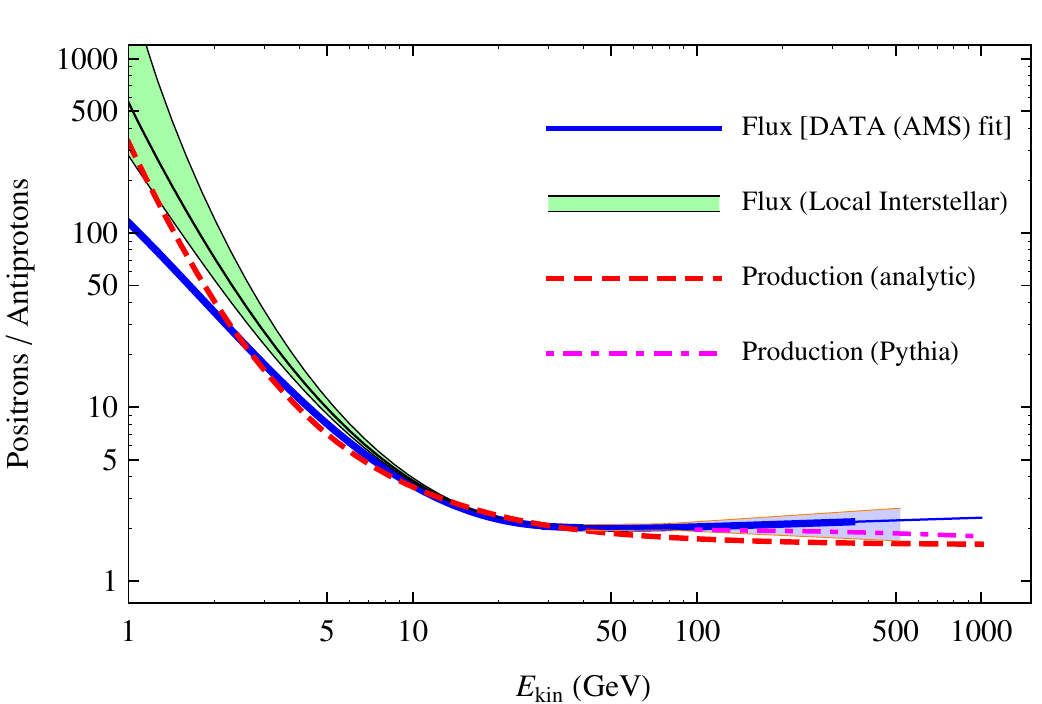}
\end{center}
\caption {\footnotesize
Ratio $e^+/\overline{p}$ as a function of the kinetic energy.
The thick line shows the ratio $\phi_{\overline{p}}(E) / \phi_{e^+} (E)$
of the fluxes observed at the Earth (the line is also shown in Fig.~\ref{fig:ratio-ep}).
The ratio of the fluxes in interstellar space is shown as a shaded
are, and was calculated correcting for the solar modulation effects
using three different values of the solar modulation parameter
($V = 0.4$, 0.6 and 0.9~GV).
The dashed line is the calculated ratio of the production rates
of positrons and antiprotons in the solar neighborhood.
For $E > 100$~GeV also the results of calculation performed
using the Pythia Monte Carlo code is shown.
\label{fig:ratio-anti} }
\end{figure}

\begin{figure} [hbt]
\begin{center}
\includegraphics[width=14.0cm]{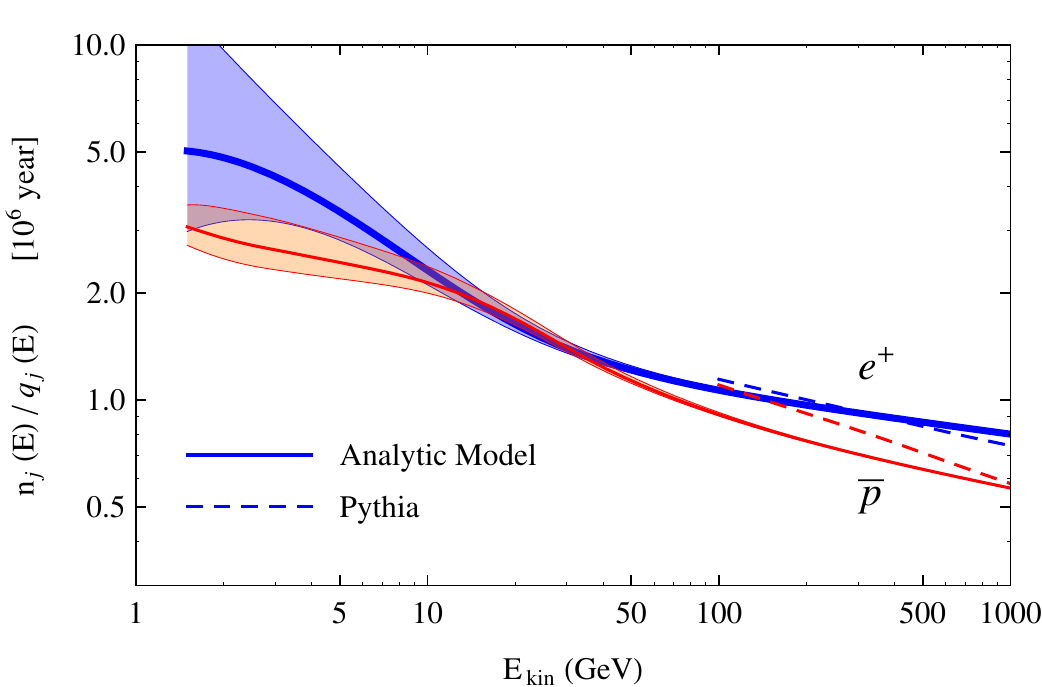}
\end{center}
\caption {\footnotesize
 Ratio $n_j(E)/q_j (E)$ (with $j = \overline{p}, e^+$)
 between the measured cosmic ray density and the production rate in the solar
 neighborhood calculated in the conventional model of secondary production.
\label{fig:time} }
\end{figure}

\begin{enumerate}
\item Uncertainties associated to the description of the solar modulations
 are significant at low energy, but vanish for $E \gtrsim 20$--30~GeV, when
the calculations performed with different value of the 
potential $V$ become approximately equal.

\item For $E \gtrsim 30$~GeV the spectra of $\overline{p}$ and $e^+$ 
take a simple power law form ($q_j (E) \propto E^{-\alpha_j}$) with approximately the same exponent.

\item The exponents $\alpha_{e^+}$ and $\alpha_{\overline{p}}$ of the positron and 
antiproton production rates have a value close 
to the spectral index of the primary proton spectrum:
\begin{equation}
\alpha_{\overline{p}} \simeq \alpha_{e^+} \approx \gamma_p ~.
\label{eq:exponents} 
\end{equation}
This is an expected and well understood result:
the energy distributions of secondaries generated by a power law spectrum 
of primary particles, in good approximation
and sufficiently far from threshold effects,
have a power law form with the same spectral index 
(see also discussion below in the appendix~\ref{sec:analytic}).
Note however that equation (\ref{eq:exponents}) is not an exact result.
This is because the primary spectrum is not a simple power law
(and therefore $\gamma_p$ is not exactly constant),
and scaling violations in the hadronic cross sections also
introduce small but non negligible (and model dependent) corrections.

\item The ratio $q_{e^+} (E) /q_{\overline{p}} (E)$ 
of the positrons and antiprotons injection rates
is shown in Fig.~\ref{fig:ratio-anti} plotted as a function of the kinetic energy. 
For small energy the injection rate of positrons 
is much larger than for antiprotons
($q_{e^+} (E) /q_{\overline{p}} (E) \simeq 500$ at $E \simeq 1$~GeV).
The ratio decreases monotonically with $E$ 
and reaches an asymptotic value of order~2 for $E \gtrsim 20$--30~GeV.
This behavior can be easily understood qualitatively as a
the consequence of simple kinematical effects.
The production of antiprotons at low energy is suppressed 
because the creation of a nucleon/antinucleon pair has a relatively
high energy threshold
(the minimum kinetic energy of the projectile proton is $E_{0,{\rm min}} = 6~m_p$). 
In addition the creation of antinucleons at rest or
moving backward in the laboratory frame
is kinematically forbidden for any projectile energy $E_0$.
The importance of these kinematical effects becomes progressively less important
with increasing energy, and the ratio $e^+/p$ converges
to an approximately constant value.
\begin{equation}
 \left .\frac{q_{e^+}(E)}{q_{\overline{p}}(E)}
 \right |_{E \gtrsim 30~{\rm GeV}} \approx 1.8 \div 2.1
\label{eq:ratio-2}
\end{equation}
It can be interesting to  have a  simple   analytic   calculation  of 
the $e^+/\overline{p}$  ratio at production in the limit of high energy
(when the ratio is approximately constant).  Such a simple calculation
allows to  discuss an estimate  of the different systematic  uncertainties.
Such a simple  calculation is outlined  in appendix~\ref{sec:analytic}.

\item Figure~\ref{fig:ratio-anti} also shows the 
ratio $n_{e^+} (E)/n_{\overline{p}}(E)$ of the densities 
of antiparticles in the local interstellar medium
(estimated deconvolving the effects of solar modulations).
The two ratios have energy dependences that are remarkably
similar. In fact, taking into account the systematic uncertainties, the 
two ratios are consistent with being equal in the entire energy range:

These results can be summarized with the statement:
\begin{equation}
 \frac{\phi_{e^+}^{\rm loc} (E)}{\phi_{\overline{p}}^{\rm loc}(E)}
 \approx
\frac{q_{e^+}^{\rm loc} (E)}{q_{\overline{p}^{\rm loc}}(E)} ~.
\label{eq:ratio-surprise}
\end{equation}
an unexpected and intriguing result. 

\item
  The ratio between the local number density and the local injection rate
  for cosmic rays   of type $j$
  gives an (energy dependent)  characteristic time:
\begin{equation}
\tau_j (E) =  \frac{n_{j}^{\rm loc} (E)}{q_{j}^{\rm loc}(E)}
 \approx
\label{eq:timej} 
\end{equation}
  Equation (\ref{eq:ratio-surprise}) implies that
  the characteristic  times  for positrons and  antiprotons are
  approximately equal  as shown in Fig.~\ref{fig:time}.
  Inspection of this figure shows that the characteristic  times
  are of order 10$^6$~years, and have 
  a weak but not trivial energy dependence, most clearly in the low energy
  region ($E \lesssim 20$--30~GeV).
  For  $E \gtrsim 50$~GeV the energy dependence of the
  $e^+$ and $\overline{p}$  characteristic times
  is weaker,
  and a power law description of the numerical results
  suggests an approximate dependence $\propto E^{0.15 \pm 0.07}$,
  where the error is a rough
  estimate of the systematic uncertainties associated
  to the shape of the primary  flux and of the modeling
  of hadronic interactions.

 The quantities $\tau_j (E)$ can be related to the residence time of particles
 in the Galaxy, but the relation depends on the effective volume of the production
 and containement of the particles.
\end{enumerate}

The result of equation (\ref{eq:ratio-surprise}) that the ratio of
the production rates of positrons and antiprotons in the solar neighborhood,
is (within systematic uncertainties)
equal to the ratio of the fluxes of $e^+$ and $\overline{p}$ is a
result that naturally suggests that the local production of
antiparticles is representative of the the production rate in the entire
Galaxy (or in an important part of the Galaxy), and that the $e^+/\overline{p}$ ratio
at production is preserved by propagations.
It is of course also possible that equation (\ref{eq:ratio-surprise})
is only a ``coincidence'', a numerical accident of no physical significance,
and that the positron and antiproton fluxes are generated
by distinct source mechanisms.
The implications of these two possibilities will be discussed in the following.

\section{Energy losses}
\label{sec:losses}

The rates of energy loss for relativistic $e^\pm$ and $p$ ($\overline{p}$) 
propagating in interstellar space differ by many orders of magnitude.
The dominant mechanisms of energy losses are synchrotron emission and
Compton scattering (where the target is formed by the radiation fields
present in space: CMBR, infrared radiation and starlight).
In both cases the rate energy losses depends on the mass and energy
of the particles approximately as $|dE/dt| \propto E^2/m^4$, so that 
the effect is completely negigible for $p$ and $\overline{p}$, but is
potentially important for $e^\pm$.

The rate of energy loss of $e^\mp$ due to synchrotron radiation is:
\begin{equation}
-\left . \frac{dE}{dt} \right |_{\rm syn} = 
\frac{4}{3} \; \sigma_{\rm Th} \, c \; \rho_B \; \frac{E^2}{m_e^2} 
\label{eq:syn}
\end{equation}
where $E$ is the particle energy, 
$\rho_B = B^2/(8 \pi)$ is the energy density stored in the magnetic field,
and $\sigma_{\rm Th}$ is the Thomson cross section.

The energy loss for Compton scattering depends
on the density and energy distribution of the photons that form 
the target radiation field.
In a reasonably good approximation the energy loss of $e^\mp$ for
Compton scattering is given by an expression very similar to
(\ref{eq:syn}): 
\begin{equation}
-\left . \frac{dE}{dt} \right |_{{\rm IC}} \simeq 
\frac{4}{3} \; \sigma_{\rm Th} \, c 
\; \rho_\gamma^* (E) 
\; \frac{E^2}{m_e^2} 
\label{eq:IC}
\end{equation}
where $\rho_\gamma^*(E)$ is the energy density in target photons
that have energy below the maximum value $\varepsilon_{\rm max} \simeq 4\, m_e^2/E$.
This constraint insures that the $e \gamma$ scatterings are in the so
called Thomson regime. Interactions with higher energy photons are
in the Klein--Nishina regime, where the $e\gamma$ cross section is suppressed.

Combining the energy losses for synchrotron emission and Compton scattering,
the characteristic time for energy loss for $e^\mp$ is:
\begin{eqnarray}
 T_{\rm loss} (E) & = & \frac{E}{|dE/dt|} \simeq
 \frac{3 \, m_e^2}{4 \, c\, \sigma_{\rm Th} \, \langle \rho_B + \rho_\gamma^*(E) \rangle \; E} 
\nonumber \\[0.2cm]
& \simeq &
310.8 ~
\left [ \frac{\rm GeV}{E} \right ]~
\left [ \frac{{\rm eV}\;{\rm cm}^{-3}}
 {\left \langle \rho_B + \rho_\gamma^* (E) \right \rangle}
\right ]
 ~{\rm Myr}~.
\label{eq:loss-num}
\end{eqnarray}
In this equation the average $\langle \rho_B + \rho_\gamma^*(E) \rangle$ is performed along the
trajectory of the particle.

The characteristic time for energy loss scales $\propto E^{-1}$, and depends
on the average energy density $\langle \rho_B + \rho_\gamma^*(E) \rangle$.
To estimate this energy density one can observe that the energy density of the
magnetic field $\rho_B = B^2/(8 \pi)$ scales as the square of the magnetic field,
and for $B \simeq 3~\mu$G (the typical strength of the magnetic field
in the Galactic plane) has the value $\rho_B \simeq 0.22$~eV/cm$^3$.
The energy density of the radation field can be decomposed in the sum of three
main components: CMBR, dust emitted radiation and starlight.
The 2.7~Kelvin Cosmic Microwave Background Radiation (CMBR) fills
homogeneously all space with the energy density $\rho_{\rm CMBR} \simeq 0.26$~eV/cm$^3$.
CMBR photons have the average energy 
$\langle \varepsilon \rangle \simeq 6.3 \times 10^{-4}$~eV,
and therefore this component is
effective as a target for Compton scattering
up to very high energy ($E \lesssim 400$~TeV).
The dust emitted infrared radiation in the solar neighborhood has an energy density
of order $\rho_\gamma \simeq 0.25$~eV/cm$^3$, formed by infrared
photons with an average energy of order 0.01~eV,
and is effective for $E \lesssim 30$~TeV.
Finally starlight in the solar neighborhood has an energy density of order
$\rho_\gamma \simeq 0.5$~eV/cm$^3$, formed by photons with an average energy
of 1~eV, and is effective for $E \lesssim 300$~GeV.
One can conclude that the energy loss time of an $e^\mp$ with energy
of order 1~TeV is of order 0.6~Myr if the particles remain close to the
galactic plane, but can also be longer (of order of 1.2~Myr)
if the particle confinement volume is much larger than the galactic disk.

\section{Formation of the Cosmic Ray fluxes}
\label{sec:formation}
The formation of the Galactic cosmic ray fluxes
can be naturally decomposed into two steps: ``release'' and ``propagation''.
In the first step, relativistic charged particles
are injected or ``released'' in interstellar space. In the second step
the particles propagate in the Galaxy from the release point
to the observation point.

Assuming the validity of this decomposition, the relation
between $n_j (E, \vec{x}, t)$ 
(the density of cosmic rays of type $j$ and energy $E$ present
at time $t$ at the point $\vec{x}$) 
and $q_j(E_i, \vec{x}_i, t_i)$ (the rate of release per unit volume
of cosmic rays of type $j$ and energy $E_i$ at time $t_i$ at the point $\vec{x}_i$)
can be written in general as:
\begin{equation}
n_{j} (E, \vec{x}, t) =
\int_{-\infty}^{t} dt_i ~ \int d^3x_i ~\int dE_i 
~q_{p} (E_i, \vec{x}_i, t_i) \times P_{j} (E, \vec{x}, t; ~ E_i, \vec{x}_i, t_i) ~.
\label{eq:prop-general}
\end{equation}
In this equation
$P_{j} (E, \vec{x}, t; ~ E_i, \vec{x}_i, t_i)$
is a propagation function that describes the probability
density that a particle of type $j$ that at the time $t_i$ is at the point $\vec{x}_i$
with energy $E_i$ can be found at a later time $t$ at the point $\vec{x}$
with energy $E$.
The subscript in the notation indicates that
the propagation functions will in general have different forms
for different particle types.

It is natural to divide the cosmic rays into two classes, according to the nature
of their mechanism of interstellar release. Primary particles
(such as proton, electrons, or helium nuclei) are accelerated in
astrophysical sources and then ejected into interstellar space.
Secondary particles (such as the rare nuclei Lithium, Beryllium and Boron)
are created already relativistic as final state products in the inelastic
interactions of primary cosmic rays.

Equation (\ref{eq:prop-general}) is written in a very general form, using an
absolute minimum of theoretical assumptions: it does not requires that the Galactic
CR are in a stationary state, it allows for CR in different points of the Galaxy
to have spectra of different shapes, it allows for energy losses or
reacceleration during propagation, and leaves the propagation function to have
a completely general form.
Most models for the Galactic cosmic rays have to introduce several
simplifying approximations.

Without loss of generality it is possible to define the
current total rate of release of CR of type $j$ as:
\begin{equation}
Q_j (E) = \int d^3 x ~ q_j (E, \vec{x}, t_0)~.
\end{equation}
The flux of CR of type $j$ can be written in the form:
\begin{equation}
 \phi_j (E) \approx ~\frac{\beta \, c}{4 \, \pi} \; Q_j (E) ~\frac{T_{j} (E)}{V_{\rm eff}}
\label{eq:flux-time}
\end{equation}
where $V_{\rm eff}$ is an effective Galactic volume (taken as
independent from the particle type and energy),
and $T_{j} (E)$ is an (energy dependent) quantity with the dimension of time that
has to be calculated constructing a model for CR propagation in the Galaxy.
The spectral shapes of CR at the Sun are determined by the
combination of two effects:
the properties of the sources (that determine of the release spectra),
and the properties of propagation.

\subsection{The ``standard framework'' for Cosmic Rays in the Galaxy}
\label{sec:standard}

In recent years several authors have discussed 
the problem of the formation of the Cosmic Rays spectra
in the Galaxy using a set of very similar ideas
and assumptions, that can be seen as constituting a 
common, broadly (but not universally) accepted framework for the 
formation of the cosmic ray spectra.
We will refer here to this set of ideas and assumptions 
as the ``standard framework'' for Cosmic Rays in the Galaxy and outline
below some of the most relevant points. 

\begin{itemize}
\item [(a)] 
Relativistic protons and electrons are released in interstellar space
by the astrophysical accelerators with energy spectra that,
in a broad energy range, have approximately the same shape, so that
the $e^-/p$ ratio is approximately constant.

 The spectra generated by the accelerators have the same form because the 
 particles are accelerated by a ``universal'' mechanism,
 such as Fermi Diffuse Acceleration (DSA),
 that operates equally on all relativistic charged particles, independently from the
 sign of the electric charge. The existence of such universal mechanism also
 insures that the spectra generated by different sources have
 approximately the same shape (perhaps differing only for having different
 maximum acceleration energies).
 
 The difference in mass for electrons and protons results in two important effects.
 (i) The efficiency for injection in the acceleration mechanism (when the particles
 are extracted from the tail of thermal distribution of the medium) is higher
 for protons than for electrons. This results in a (constant) $e^-/p$ ratio
 (for the release spectra) that is much smaller than unity.
 (ii) The maximum energy that a source can impart to electrons
 is expected to be smaller than the maximum energy for protons, since
 it corresponds to the energy for which
 the rates of acceleration and energy loss are equal.

\item [(b)]
 The propagation of protons and nuclei is controlled by diffusion generated
 by the irregular Galactic magnetic field. A fundamental quantity
 is the residence (or escape) time $T_{\rm esc}(E)$, that is the average time
 for a cosmic ray to diffuse out of the Galaxy. The escape time is a function
 of (the absolute value) the particle rigidity, and scales as $\propto \beta^{-1}$.

 The characteristic time for protons or nuclei
 ($T_p (E)$ or $T_A(E)$) that enters equation (\ref{eq:flux-time}) can be identified
 with the residence time.

\item [(c)]
 The rigidity dependence of the CR residence time
 can be determined comparing
 the spectral shapes of primary and secondary nuclei.
 The comparison of the spectra of Boron and Carbon suggests the
 power law dependence:
 \begin{equation}
 T_{\rm esc} (p/|Ze|) \propto \left ( \frac{p}{|Z e|} \right )^{-\delta}
\end{equation}
with $\delta \simeq 0.4$--0.5.

\item[(d)]
 The difference in spectral shape between electrons and protons
 is a propagation effect, and is the consequence of the much larger
 rate of energy loss for electrons.
 
\item[(e)]
 The $e^-/p$ ratio start falling for $E$ larger than few GeV.
 If this is the effect of different energy losses for $e^-$ and $p$,
 one has to infer that for $E \simeq 10$~GeV,
 the electron residence is longer than the energy loss time
 (estimated in equation (\ref{eq:loss-num})):
 $T_{\rm esc} (10~{\rm GeV}) \gtrsim 30$~Myr.

\item[(f)]
 The characteristic time $T_e(E)$ in equation (\ref{eq:flux-time})
 is a combination of the escape time $T_{\rm esc}(E)$ and energy loss time
 $T_{\rm loss}(E)$.
 When the loss time is much longer than the residence time
 ($T_{\rm loss}(E) \gg T_{\rm esc} (E)$) one has
 $T_e (E) \simeq T_{\rm esc} (E)$. In the opposite case
 ($T_{\rm loss}(E) \ll T_{\rm esc} (E)$) the resulting 
 $T_e(E)$ depends on the space distribution of the CR accelerators
 and the geometry of the CR confinement volume.
 For example, if the electron release and confinement volumes are identical,
 one has that $T_e(E) \simeq T_{\rm loss}(E) \propto E^{-1}$.
 If the CR electrons are generated in a disk region much thinner 
 than the confinement volume, then $T_e(E) \propto E^{-(1 + \delta)/2}$
 where $\delta$ is the exponent that describes the energy
 dependence of $T_{\rm esc} (E)$.
 More in general the energy dependence of the characteristic electron time
 can be parametrized with the power law form
 $T_e(E) \propto E^{-\delta_e}$ with an exponent $\delta_e > \delta$,

 The assumption that electrons and protons are released in interstellar space
 with the same energy distribution implies that the $e^-/p$ ratio
 measures the different energy dependence
 of the characteristic times $T_p(E) \simeq T_{\rm esc}(E)$ and $T_e(E)$:
 \begin{equation}
 \frac{\phi_{e^-} (E)}{\phi_{p} (E)}
 \simeq
 \frac{T_{e} (E)}{T_{p} (E)}
 \simeq
 \frac{T_{e} (E)}{T_{\rm esc} (E)}
 \propto E^{-(\delta_e -\delta)}~.
\end{equation}

\item[(g)]
 The spectra of CR protons and nuclei are expected to have approximately equal
 shape in different points of the Galaxy, while the shape of the 
 CR electron spectrum should become softer for points distant from
 the Galactic plane. This is because the electron accelerators are close
 to the Galactic plane, and the particles lose a non negligible amount
 of energy during propagation.
\end{itemize}

\subsection{Antiparticle fluxes in the ``standard framework''}
Using the chain of arguments outlined above it is possible to estimate
the fluxes of positrons and antiprotons generated by the conventional
mechanism of secondary production. The result is a 
predicted positron flux that is significantly softer than the observations.

It is straightforward to present the result stated above for 
the energy range $E \gtrsim 30$~GeV, where the CR spectra have power law form.
\begin{itemize}
 \item 
 The spectra of secondary positrons and antiprotons generated
 in hadronic interactions have approximately the same spectral index as
 the primary protons (and nuclei):
\begin{equation}
\alpha_{e^+} \simeq \alpha_{\overline{p}} \simeq \gamma_p
\end{equation}

\item
 The properties of propagation for $e^+$ and $e^-$ are approximately equal,
 and similarly for $p$ and $\overline{p}$.
 This implies that the production spectra of $e^+$ and $\overline{p}$ are
 distorted by the propagation in distinct ways that are
 predicted by the observations discussed above.
 
 The prediction for the positron/antiproton ratio is:
\begin{equation}
 \frac{\phi_{e^+} (E)}{\phi_{\overline{p}} (E)}
 \simeq
 \frac{\phi_{e^-} (E)}{\phi_{p} (E)}
 \simeq
 \frac{T_e(E)}{T_{\rm esc}(E)}
 \simeq E^{-(\delta_e - \delta)}
\label{eq:posi-pbar-ratio}
\end{equation}

Other robust predictions are that 
positron spectrum should be softer than the electron spectrum,
and the $\overline{p}$ should be softer than the $p$ spectrum, with
approximately equal distortions: 
\begin{equation}
 \frac{\phi_{\overline{p}} (E)}{\phi_{p} (E)}
 \simeq
 \frac{\phi_{e^+} (E)}{\phi_{e^-} (E)}
 \simeq
 E^{-\delta}
\label{eq:posi-ele-ratio}
\end{equation}
\end{itemize}

The predictions presented in
equation (\ref{eq:posi-pbar-ratio}) and (\ref{eq:posi-ele-ratio})
are not supported by the data.
The most spectacular discrepancy is for the positron flux,
that is predicted to be softer than the electron one while in the
data the $e^+$ spectrum is {\em harder},
in direct conflict with the prediction.

In the standard framework for CR in the Galaxy, the hard spectrum of positrons
is explained introducing a new source of relativistic positrons 
(such as direct acceleration in Pulsars or the annihilation or decay of Dark Matter).
The shape of the spectrum generated by this new positron
source is harder than the spectrum of positrons generated by the
conventional mechanism so that, including the softening induced by the
propagation, one obtains the observed spectrum.

The new positron source cannot generate a significant spectrum
of antiprotons with a spectrum of similar shape,
because in the ``standard framework'' we are discussing here
such a new source would result
in an antiproton spectrum harder than the positron one, and harder
than the observations.

The conclusion is that the observed positron and antiproton fluxes
must be generated by distinct mechanisms, and are completely unrelated.
The fact that they have the same spectral index ($\gamma_{e^+} \simeq \gamma_{\overline{p}}$
for $E > 30$~GeV) is just
an accident, because these indices emerge in completely different ways.
For positrons one has
\begin{equation}
\gamma_{e^+} \simeq \alpha_{e^+} + \delta_e
\end{equation}
with $\alpha_{e^+}$ the spectral index generated by the new source, and $\delta_e$
related to the propagation effect for $e^\pm$, while 
for antiprotons
\begin{equation}
\gamma_{\overline{p}} \simeq \alpha_{\overline{p}} + \delta
\end{equation}
with $\alpha_{\overline{p}}$ the spectral index that emerge
from the conventional mechanism, and $\delta$ that describes propagation effects
for $p(\overline{p})$.

The approximate equality of the spectral indices of positrons and antiprotons
is already intriguing, but perhaps even more surprising is the fact that 
two unrelated mechanisms generate spectra of approximately the same
absolute value, in fact that the 
{\em observed} positron/antiproton ratio (approximately $2.04 \pm 0.04$)
is equal (within systematic uncertainties)
to the ratio at {\em production} calculated with the conventional mechanism.
The last result can in fact be extended to the broader 
energy range $E \in [1,350]$~GeV, as shown in equation
(\ref{eq:ratio-surprise}), strongly suggesting that we are not in front to
a numerical coincidence and that a physical explanation is required.

\section{Alternative Framework}
\label{sec:alternative}

In this section we will take the point of view,
that the result of equation (\ref{eq:ratio-surprise})
is not a numerical accident, but is an essential indication
that points to a common origin for the observed fluxes
of $e^+$ and $\overline{p}$, and discuss if this point of view
is viable, and what are its implications.

Equation (\ref{eq:ratio-surprise})
shows that, taking into account statistical and systematic uncertainties,
the ratio of the observed fluxes of positrons and antiprotons,
is equal to the ratio for the production rate of antiparticles
in the solar neighborhood, where the mechanism of production of $e^+$ and $\overline{p}$
is the creation of secondaries in the hadronic interactions of primary cosmic rays.
If the result is not a simple numerical coincidence, it suggests the following
points:
\begin{itemize}
\item[(i)]
Secondary production in hadronic interactions is indeed the
source of the CR positrons and antiprotons.

\item[(ii)]
 The energy dependence of the ``local'' (i.e. solar neighborhood)
 production of antiparticles is representative for the entire (or most of the)
 production of antiparticles, at least for our space--time point of observation.

\item[(iii)]
 Propagation effects must not distort the energy dependence of the $e^+/\overline{p}$
 ratio at production. From this one can deduce two results.
 \begin{itemize}
 \item[(a)] Positrons and antiprotons must propagate in approximately the same way.
 \item[(b)] The energy of the particles must remain approximately constant during propagation. 
 This is because the $e^+$ and $\overline{p}$ production spectra
 have different shapes,
 and a significant variation of the energy of the particles during propagation
 would result in distinct spectral distortions for the two particle types 
 and change the energy dependence of the ratio.
\end{itemize}
\end{itemize}

The general expressions for the formation of the spectra of
cosmic rays has been given in equation (\ref{eq:prop-general}).
It is possible to express more formally the points just listed above
using the same notation.
Point (i) and (ii) suggest that the production rates of positrons and
antiprotons have the factorized form:
\begin{equation}
 q_{e^+(\overline{p})} (E, \vec{x}, t) \simeq
 q_{e^+(\overline{p})}^{\rm loc} (E) ~ F_q (\vec{x}, t) 
\label{eq:prop1}
\end{equation}
so that the energy dependence of the local production is valid in the 
region of space and time where the production of antiparticles is important.
The function $F_q(\vec{x},t)$
that describes the space and time depedence of
the antiparticles production must be equal for positrons and antiprotons.
This is because, the observed fluxes of $e^+$ and $\overline{p}$ are formed summing
contributions produced in different points and at different times. 
To insure that the ratio of the observed fluxes is equal to the ratio
at production, the space--time distributions
of $q_{e^+}$ and $q_{\overline{p}}$ must be equal.

In addition, as stated in point (iii), the propagation of positrons
and antiproton must be (approximately) equal, and leave
the energy of the antiparticles (approximately) constant. This can be obtained
when the propagation functions for positrons and antiprotons in
equation (\ref{eq:prop-general}) take the form:
 \begin{equation}
 P_{e^+ ( \overline{p})} (E, \vec{x}_\odot, t_0; ~E_i, \vec{x}_i, t_i) \simeq
 \delta [E - E_i] ~ P (E, \vec{x}_\odot, t_0; ~ \vec{x}_i, t_i)~
\label{eq:prop2}
\end{equation}
 where $\vec{x}_\odot$ is the position of the solar system in the Galaxy, $t_0$ the time now,
 and the function $P$ is equal for positrons and antiprotons.
 
 Inserting equations (\ref{eq:prop1}) and (\ref{eq:prop2}) in
 the general expression (\ref{eq:prop-general}) for propagation one obtains, for the density
 of positrons and antiprotons at the present time in the vicinity of the Sun
 the expressions:
\begin{eqnarray}
 n_{e^+(\overline{p})}^{\rm loc} (E)
& = & 
q_{e^+(\overline{p})}^{\rm loc} (E)~ \left [
\int_{-\infty}^{t_0} dt_i ~ \int d^3x_i 
~F_{q} (E, \vec{x}_i, t_i)
\times P (E, \vec{x}_\odot,t_0; ~ \vec{x}_i, t_i) \right] \\[0.2 cm]
 & = & q_{e^+(\overline{p})}^{\rm loc} (E)~ \tau(E)
\end{eqnarray} 
The factor in square parenthesis in the equation above has the dimension
of time, and is common for positrons and antiprotons
so that equation (\ref{eq:ratio-surprise}) is satisfied.

Is it possible to construct a model where the formation of the spectra of positrons
and antiproton happens with the properties listed above ?
The most critical problem is the requirement that the propagation of
positrons and antiproton is (at least approximately) equal.
The {\em rates} of energy losses for the two particle types differ by
many orders of magnitude, therefore the only way to insure that positrons
and antiprotons have the same properties in Galactic propagation is that
the residence time of the particles in the Galaxy is sufficiently short, so that
the total energy loss $\Delta E \simeq |dE/dt| \; T_{\rm esc} (E)$
suffered by positrons during propagation remains negligible
($\Delta E \ll E$). This requirement can also be expressed in the form: 
\begin{equation}
T_{\rm esc} (E) < T_{\rm loss} (E) \lesssim 1.2 ~
\left [ \frac{1}{E_{\rm TeV}} \right ]~
\left [ \frac{0.26 ~{\rm eV}\,{\rm cm}^{-3}}
 {\left \langle \rho_B + \rho_\gamma^* (E) \right \rangle}
\right ] ~{\rm Myr}
\label{eq:time-upper-bound}
\end{equation}
where we have made use of equation (\ref{eq:loss-num}).
The numerical upper limit on the residence time
is obtained setting the energy density
$\langle \rho_B + \rho_\gamma^* (E) \rangle$ equal to its minimum value,
that is the energy density of the CMBR
that permeates uniformly space. This is possible 
if the confinement volume of the particles is very large, so that the contributions
to the energy density of the magnetic field and of radiation generated in the Galaxy
(that are concentrated near the Galactic disk) become negligible.

\begin{figure} [hbt]
\begin{center}
\includegraphics[width=14.0cm]{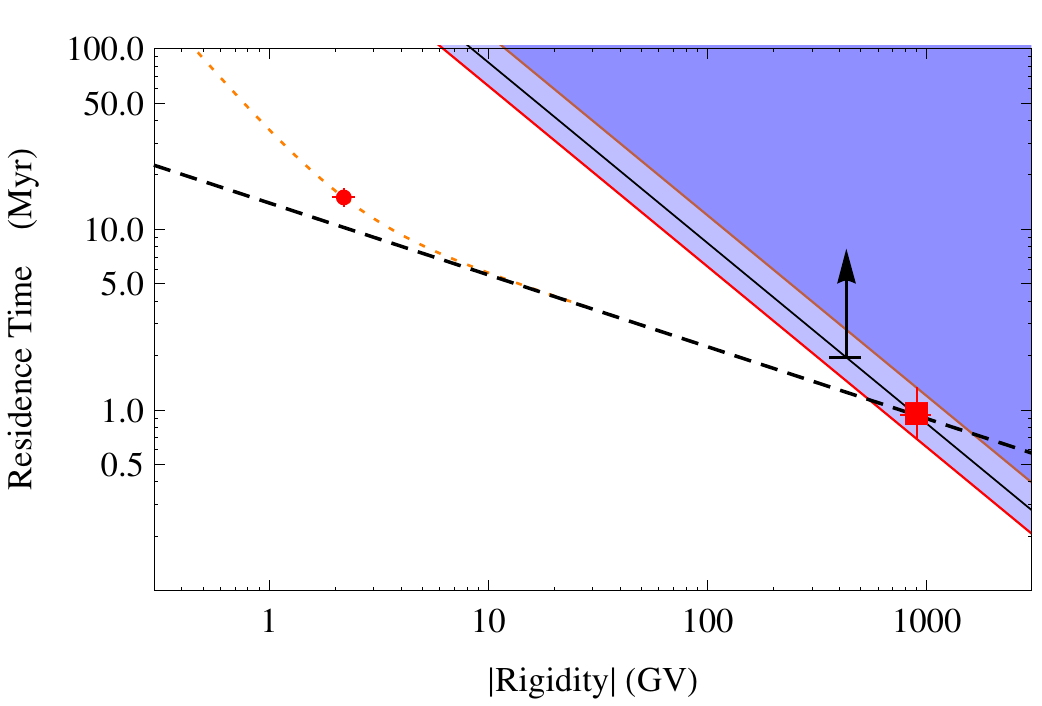}
\end{center}
\caption {\footnotesize
Contraints on the residence time of cosmic rays in the Galaxy.
The solid lines are lower limits on $T_{\rm res}$ obtained assuming that
positrons with $E \lesssim 430$~GeV suffer neglible energy loss
during their residence in the Galaxy. The different lines assume different
confinement volumes and therefore different values for the
average energy loss rate. The arrow corresponds to the 
highest energy for which a measurement of the $e^+$ flux is available.
The point at low rigidity is the estimate on the confinement time
$T_{\rm res} = 15.0 \pm 1.6$~Myr obtained by the CRIS collaboration
 \cite{beryllium3} interpreting their measurement of the $^{10}$Be/$^9$Be ratio. 
The point at rigidity 900~GeV is estimated assuming that the break observed
by HESS in the $(e^+ + e^-)$ spectrum corresponds to the critical energy
where the residence and energy loss time are equal. The dotted line 
that connects the two points is
given by expression (\ref{eq:tres-parametrization})
with the parameters $T_0 \simeq 14.0\pm 0.4$~Myr and $\delta\simeq 0.40 \pm 0.04$.
 \label{fig:timebounds} }
\end{figure}

The upper bound on the CR residence time 
in equation (\ref{eq:time-upper-bound}) is
shown graphically in Fig.~\ref{fig:timebounds}.
Is such a bound compatible with existing observations? 
The most direct and reliable method to estimate the residence time of 
cosmic rays in the Galaxy is based on the study of 
the ratio of the beryllium isotopes ${}^{10}$Be and 
${}^{9}$Be. Both isotopes are formed as secondaries
in the fragmentation of higher mass nuclei,
but the beryllium--10 is unstable
with an half--life of $1.51\pm 0.04$~Myr \cite{Tilley:2004zz} and 
acts as a cosmic--clock to measure the time elapsed from the nucleus
creation \cite{beryllium1,beryllium2,beryllium3}.
The Cosmic Ray Isotope Spectrometer (CRIS) collaboration,
using an instrument aboard the Advanced Composition Explorer (ACE) spacecraft
has measured the ratio ${}^{10}$Be/${}^{9}$Be in the energy range
$E_0 \simeq 70$--145~MeV/nucleon (with $E_0$ the kinetic energy per nucleon).
The ratio is of order 0.11--0.12, and since the production rate of the two isotopes
in the fragmentation of higher mass nuclei is approximately equal, this
indicates that approximately 90\% of the Beryllium--10 nuclei have decayed.
The CRIS collaboration, using a
a ``Leaky Box'' theoretical framework has then estimated 
a residence time $T_{\rm esc} = 15.0 \pm 1.6$~Myr for the unstable nuclei.

In order to extrapolate these results to $e^\pm$ in the energy range of
several hundred GeV, one can make the ansatz that the residence time
depends on the rigidity and velocity of the particles with the
(2--parameter) functional form:
\begin{equation}
T_{\rm res} \approx \frac{T_0}{\beta} \; \left ( \frac{p}{p_0 \; |Ze|}\right )^{-\delta_T} ~
\label{eq:tres-parametrization}
\end{equation}
($p_0$ is an arbitrary scale that without loss of generality can be set to 1~GeV).
The constraint that the residence time is shorter than $T_{\rm loss} (E)$ for
$E \lesssim E^*$ can then be used to set limits on the parameters $\delta_T$ and $T_0$.

An intriguing possibility is to interpret the spectral break
observed by the Atmospheric Cherenkov telescopes as the transition
from the regime where the residence time for $e^\pm$ is shorter than
the loss time, to the regime where the opposite is true.
Defining the critical energy $E^*$ as the energy where the
residence and loss time for $e^\mp$ are equal:
\begin{equation}
T_{\rm loss} (E^*) \simeq T_{\rm res} (E^*)~
\end{equation}
one expects to observe a spectral feature in correspondence of $E^*$ with a
$\Delta \gamma$ that can be as large as unity.

Tentatively identifying the break energy observed by HESS, MAGIC and VERITAS
with the critical energy $E^*$,
one can estimate the residence time for $e^\mp$ at the break energy,
as it is equal to the loss time. The result (using the HESS estimate for
the energy of the break energy): 
\begin{equation}\
T_{\rm res} (900~{\rm GeV}) \simeq T_{\rm loss} (900~{\rm GeV}) ~
\simeq 1.0 \pm 0.4~{\rm Myr}
\label{eq:tloss-hess}
\end{equation}
where the error was estimated taking into account a 20\% uncertainty
in the energy scale, and a range of possible values
for $\langle \rho_B + \rho_\gamma^* \rangle$ in equation (\ref{eq:loss-num}).

Combining the two measurements of the CR residence time obtained at
different rigidities, performed by CRIS using the beryllium ratio \cite{beryllium3},
and by the Air Cherenkov telescopes, and comparing to
equation (\ref{eq:tres-parametrization}), one can estimate the two parameters
$T_0$ and $\delta_T$ obtaining the results:
$T_0 \simeq 14.0\pm 0.4$~Myr and $\delta\simeq 0.40 \pm 0.04$.
The rigidity dependence of the residence time obtained in this estimate is actually
consistent with the indications obtained from the ratio
$q_{j}^{\rm loc}(E)/n_j^{\rm loc}(E)$ discussed in section~\ref{sec:hadronic} and shown
in Fig.~\ref{fig:time}.

The short CR residence time estimated above is also an important constraint in
construction of a model of the Galactic magnetic field, since the
outflow of the CR particles from the Galaxy (that is inversely proportional
to the residence time) must generate, angular anisotropies
smaller than the experimental upper limits. 
Simple diffusive models (see for example \cite{Hillas:2005cs}) 
overpredict the dipole amplitude of the CR angular distribution, and a shorter
lifetime exacerbates this problem. It is therefore necessary to
go beyond these simple models \cite{Mertsch:2014cua,lipari-anisotropy}.

The tentative conclusion we obtain from the discussion
outlined above is that an interpretation of the positron and
antiproton flux as entirely of secondary origin is in fact viable, and also
offers a simple and attractive interpretation for the softening
of the $e^\pm$ spectrum observed by the Cherenkov telescopes.

The hypothesis  that the positron and anti--proton fluxes  are of secondary
origin has  been explored in the past by several other authors see for example
\cite{Cowsik:2010zz,Cowsik:2013woa,Cowsik:2015yra,Katz:2009yd,
  Blum:2013zsa,Ahlen:2014ica}).
An important  difference of the  present paper  with respect to
previous  works  is  that here the discussion on the anti--particles  fluxes
is  constructed without making use of the data on secondary nuclei,
a topic that will be addressed in the next section.

In the present paper  we  do not have the space for
a critical review of the  the work of  the authors  that have
discussed  the  anti--particles  in cosmic rays as  secondary products.
We would like however to make a brief comment on the work 
of Blum, Katz and Waxman (BKW) \cite{Katz:2009yd,Blum:2013zsa}.
These authors conclude that both positrons and anti--protons are
of secondary origin,   but also rgue that
energy losses play a significant role in  $e^+$ propagation.
BKW introduce an energy loss suppression factor $f_{s, e^+} (E)$, 
defined as the ratio between the observed positron flux and the theoretical
flux calculated neglecting energy losses, and estimate
a value $f_{s, e^+} (E) \approx 1/3$ at $E \simeq 20$~GeV.
According to BKW the quantity $f_{s, e^+} (E)$ 
{\em grows} with energy, approaching the maximum possible value
of unity for $E \simeq 300$~GeV.
This means that the importance of the energy loss effects in shaping
the positron spectrum decreases at higher energy . This behavior
is surprising since the $e^\pm$  energy losses grow approximately quadratically with $E$. 
A factor $f_{s,e^+} (E)$ that grows with energy
requires that positrons of different $E$ remain confined
in regions of space that are not identical and contain  on average
a different magnetic field and a different energy density in radiation
(see Eq.~(\ref{eq:loss-num})) to reduce the energy losses for particles  of
higher $E$.
It is far from easy to construct a satisfactory
model to explain the energy dependence of $f_{s,e^+} (E)$ along these lines.

In contrast in this work we find find that  the comparison of the
positron and anti--proton spectra suggests  that the 
$e^\pm$ energy losses effects are negligible in the entire energy range
where data is available, and therefore, using the notation of BKW,
that $f_{s,e^+} (E) \simeq 1$  for all energies  below the critical energy
$E^*$ that has been tentatively identified as the energy
(of order 900~GeV)  where the Cherenkov telescopes observe a spectral  break
in the  ($e^- + e^+$) spectrum.
The origin of the  discrepancy between this work  and the results
of BKW  can likely be attributed to a  more precise modeling of inelastic
hadronic interactions in this work.

\section{Secondary Nuclei} 
\label{sec:secondary-nuclei}
A cornerstone of the theoretical studies  based
on the ``standard framework''  for the propagation
of Galactic cosmic rays is the interpretation of the
measurements of ``secondary  nuclei'' such as Lithium, Beryllium and Boron.
These nuclei are very rare in ordinary matter,
but they are relatively abundant in cosmic rays, because
they can be created, already relativistic in the Galaxy rest frame,
as secondary products in the fragmentation
of higher mass nuclei (mostly Carbon and Oxygen).

The ratio between the fluxes of secondary and primary  nuclei
can then be  interpreted,   using fragmentation cross sections measured
in laboratory, and correcting for absorption effects, 
to estimate the average, rigidity dependent column density
$X(|p|/Z)$ crossed by the relativistic nuclei.

The best measured quantity to perform this study
is the Boron/Carbon  (B/C) ratio, and several authors
\cite{Engelmann:1990zz,deNolfo:2006qj,Ahn:2008my,Obermeier:2012vg,Adriani:2014xoa}
have measured and interpreted this ratio.
The  estimated  average column density at a rigidity of few GV is of order of
$\approx 5$~g/cm$^2$,   and decreases with a  power law
dependence: $X(|p|/Z) \propto (|p|/Z)^{-\delta}$, 
with an exponent of order $\delta \simeq 0.3$--0.5.

In the ``standard framework'' for CR Galactic propagation
one makes two additional assumptions:
\begin{itemize}
\item [(i)]   The column  density  estimated from the B/C ratio,
  applies not only to the nuclei that enter the Boron production process,
  but also to all cosmic rays species, including
  protons and helium nuclei.
\item[(ii)] The column  density  $X$  crossed by cosmic rays
  is integrated during their propagation in
  interstellar space, and therefore it is proportional to
  the particles Galactic residence time  $T$:
  \begin{equation}
  X  \simeq  \langle \rho \rangle \; T
 \label{eq:x-t}
  \end{equation}
  (where $\langle \rho \rangle$ is the density of the  interstellar
  medium averaged over the particles trajectories).
\end{itemize}
Using these assumptions it is possible to
estimate the spectral shape of the primary cosmic rays as they are
released in interstellar space by the Galactic sources, and
to construct predictions  for the spectra of
secondary antiparticles ($e^+$ and $\overline{p}$)
that are created in the inelastic interactions  of
(mostly protons and helium nuclei) primary cosmic rays.

The  spectrum of secondary positrons constructed as outlined above
is very soft,  in strong  disagreement with  the observations.
The  softness of the $e^+$ spectrum is the consequence of the fact that
the  estimated  CR residence  time (that also  depends on the estimate
of $\langle \rho \rangle$, and therefore on the CR confinement volume)
is  sufficiently long so that the energy losses  for $e^\pm$  are important. 
As already discussed, this conflict between predictions and observations
is  commonly solved introducing a new source of relativistic  positrons.

The predictions of the spectrum of high energy antiprotons
constructed starting from  the B/C  ratio,  are closer  to the observations,
but also in this case there is significant tension (if not in open conflict)
between  predictions and data. 
In fact, all such predictions of the secondary antiprotons spectrum
constructed before the release of the AMS02 data, have 
obtained  results  significantly softer than for protons. 
For example, in the work of  Donato et al.  \cite{Donato:2008jk}
(inspecting figure~3) the  $\overline{p}/p$ ratio
for $E \gtrsim 50$~GeV   has an energy dependence  well described by
a power law ($\propto  E^{-\alpha}$)  with an  exponent   $\alpha$
in the range $\alpha  \approx  0.49 \pm 0.05$.
In the work of Trotta et al. \cite{Trotta:2010mx}
the energy dependence $\overline{p}/p$ ratio
is  accurately described (in the same  energy range)
as a power law  with exponent $\alpha \simeq 0.30$.
These results can  be easily  understood  qualitatively
observing that in these calculations the spectral index   of
the  $\overline{p}$ spectrum,  in  first approximation, takes the value:
\begin{equation}
\gamma_{\overline{p}} \simeq  \gamma_p + \delta
\end{equation}
where  $\gamma_p$ is the spectral index of the proton flux,
and $\delta$ is the exponent  of the power law  rigidity dependence
of the B/C  ratio, that   (in the  framework  in consideration)
also  give  the  rigidity dependence of the cosmic ray  confinement time.
In contrast the measurements performed  by AMS02  show  spectra
of $p$ and $\overline{p}$  that have approximately the same
energy dependence.

After the publication of the  AMS02 results,   several
authors  (for example \cite{Giesen:2015ufa,Evoli:2015vaa})
have  presented revisions  of their calculations of
of the antiproton flux that also  include  estimates
of the systematic  uncertainties  in the calculation
(the  dominant ones are the description of the hardening of the cosmic ray spectra
at high rigidity, and the modeling of  antiproton production
in inelastic hadronic interactions).
These authors argue  that,  taking into account for  these uncertainties,
their models (based on  the  ``standard  framework'')  can be
reconciled with the AMS02 data.
It should be noted however, that also in  these ``a posteriori''  revised  calculations,
the central  value of the prediction for the  $\overline{p}/p$ ratio  
falls  with energy.
In the work of Giesen et al \cite{Giesen:2015ufa}  the   energy dependence
of the ``best fit'' prediction   for the  $\overline{p}/p$ ratio 
(in the range $E \gtrsim 50$~GeV)
has a power law  form  ($\propto  E^{-\alpha}$)
with an exponent $\alpha \simeq 0.28$.
In  the calculation of Evoli et al. \cite{Evoli:2015vaa}  the
ratio falls with energy  with an exponent $\alpha \simeq 0.21$.
A detailed  discussion of this important problem is beyond the scope of this
work, but we can observe
that while the inclusion of the systematic  uncertainties
does mitigate the significance of a possible  discrepancy
between data and  prediction,  in our view,
significant tension remains.

In the present work we are taking a completely different approach,
where the origin of the fluxes of positrons and antiprotons in
cosmic rays is  studied comparing the two  antiparticles spectra  with each other
and with the spectra of the parent  particles (protons and  helium nuclei),
without any input from the observations of secondary nuclei.
As discussed above,  this  study does suggest a common secondary origin
for antiparticles.
In this approach the ratio of the observed spectra and  the spectra
generated in  the inelastic collisions of the primary particles
allows to  estimate (see Eq.~(\ref{eq:timej})) 
an energy (or rigidity) dependent characteristic time
for positrons and antiprotons that can be compared
to the column density infered from the B/C ratio,
effectively reversing the  method  used in
the studies that start with a discussion of the data on secondary nuclei.

The characteristic  times for  antiprotons and  positrons obtained 
from Eq.~(\ref{eq:timej}) appears to fall less rapidly
at high energy (or rigidity) than the  confinement times estimated from
the B/C ratio.
This naturally raises the question  of how it is possible to  reconcile
these apparently conflicting  results.
A complete discussion of this  problem  is 
beyond the scope of this  paper, it is however
appropriate to include here some comments.

The most important  discrepancy is about the
prediction of the positron spectrum.
The study performed  here suggests that the energy losses of  $e^\mp$ are  negligible,
but this  is possible  only if the CR confinement time  is  sufficiently short.
The simplest way to reconcile  a short CR  Galactic residence time
with the column density estimated from the B/C  ratio  is to 
``disconnect'' the   two  quantities,  so that Eq.~(\ref{eq:x-t}) is not valid.
This ``disconnection'' is possible  if most of the column density is  integrated
inside (or in the envelope)  of  the sources.
In fact, the column density crossed by a relativistic particle
can be decomposed into the sum:  $X \simeq X_{\rm source} + X_{\rm ism}$ where 
the first term is accumulated inside or close to its source
and the second one is integrated during propagation in interstellar space.
The possibility that the  $X_{\rm source}$  contribution is 
dominant in most (or all) the rigidity range
where  data is  available has been extensively discussed
by Cowsik, Burch and Madziwa--Nussinov  
\cite{Cowsik:2010zz,Cowsik:2013woa,Cowsik:2015yra} in the framework
of the ``Nested Leaky Box Model''.

A second question is then if the hypothesis that all cosmic rays
traverse the same (rigidity dependent) column density, equal to what is
estimated from the B/C ratio, results in spectra of secondary
positrons and antiprotons that are consistent with the observations.
In fact, the assumption of the  same power law  dependence  estimated  from
the B/C ratio for the column density traversed by protons and helium nuclei
results  in antiparticle spectra that are too soft,   because the contributions
of lower (higher) energy primaries is relatively enhanced (suppressed),
and therefore on predictions  that are
inconsistent (or at least in strong tension) with the observations.
A possible solution for this problem is to  modify the
rigidity dependence  of  the column density for high $|p|/Z$.
It should in fact be noted that the predictions of the $e^+$ or  $\overline{p}$
spectrum at energy  $E$ requires to sum the contributions of all
primary particles with energy $E_0 > E$, with most of the secondary particles
created by primaries with $E_0/E$ in the range 10--100.
This implies the calculation of the spectra  of antiparticles at high  energy
requires an extrapolation of the estimates  infered from the B/C ratio. 
Cowsik, Cowsik, Burch and Madziwa--Nussinov
\cite{Cowsik:2010zz,Cowsik:2013woa,Cowsik:2015yra} 
have argues that the  power law rigidity dependence  of the  column density infered
by the Boron/Carbon ratio is only valid for 
$|p|/Z \lesssim 200$--300~GV,  while for higher rigidities  the column density
becomes approximately constant.
This behavior reflects the transition from a
(low rigidity) regime  where the column density $X$
is  dominated  by a power law dependent  $X_{\rm source}$  component,
to a (high rigidity) regime  where an approximately constant
interstellar  contribution $X_{\rm ism}$ is  dominant.
A model of this type  can  possibly generate  secondary spectra  consistent with the
observations, and  with the results obtained  here and shown in fig.~\ref{fig:time}.

It is also very desirable to confirm the results obtained with the
study of  Boron using  other secondary nuclei (in particular
Lithium and Beryllium).
Recently the AMS02  collaboration \cite{ams-days-lithium} has presented preliminary
measurements of the CR Lithium flux that show a striking hardening of the spectrum
at a rigidity of order 300~GV,  where the spectral index  
changes  with $\Delta \gamma \simeq 0.5$--0.6. This is a spectral feature
close the one  predicted by  Cowsik and collaborators,
however the recently published AMS Boron data \cite{Aguilar:2016vqr} 
do not exhibit  a similar hardening,
and appear consistent with a simple power law, in contrast to to Cowsik et al. prediction.
These results are  surprising, because  it is expected
that the Boron and Lithium spectra, having the same  origin,
should have very similar shapes, and 
a consistent interpretation of the data appears therefore problematic, 
and a more complete critical discussion on secondary nuclei
will be therefore  postponed  to wait for the  publication  of the AMS02
measurements also for Lithium and Beryllium.

It should also  be stressed  that the interpretation of   the data on Boron 
(and  of other secondary nuclei)
depends  crucially on the  measurements  (see for example \cite{Tomassetti:2015nha}),
of the  fragmentation cross sections  of light nuclei
in reactions  such as ${}^{12}{\rm C} + p \to ({}^{10}{\rm B} + p + n) + p$,
and in fact on the extrapolation to higher  energy of the available data.
An unexpected energy dependence  of the relevant fragmentation cross sections
could  have a very significant effect on the interpretation of the data.
Given the importance of this  question, it is clearly very desirable
to obtain  new experimental  measurements in a more  extended  energy range.

A  final comment  is that the hypothesis  that
the average column density traversed  by  cosmic rays is equal for all
particle  types, and in particular is equal for light nuclei
(such as Carbon and Oxygen) and for protons and Helium nuclei
(that are the main source of antiparticles)is not necessarily correct.
The hypothesis is  natural if most of the  column density is
integrated in interstellar space, since the  trajectory  of  a particle is 
essentially determined only by its magnetic rigidity.
On the other hand, if  most of the column  density
is accumulated inside of near a  source,  and if the dominant
sources  for different  particle  types do not coincide,
then it is possible that, for the same  rigidity,
particles of different  type traverse a different average  $X$.
In this case the results obtained for secondary nuclei  cannot be applied
to the prediction of the antiparticle spectra.

To conclude  this  brief discussion  on secondary nuclei,
we can note that this  important  question certainly  deserves a more extended
and detailed discussion that will be postponed to a future paper.
Given the complexity  of this problem, we find that the approach  followed
in the present work, to tentatively explore the hypothesis  of a secondary origin
of  antiparticles in cosmic rays without including an interpretation
of the data  on secondary nuclei has its merits.  Eventually of course
a successful  theory should be able to give a satisfactory  explanation for 
all CR data. 

\section{Outlook}
\label{sec:outlook}
In this work we have shown and discussed the existence of some surprising
properties of CR positrons and antiprotons.
The ratio $e^+/\overline{p}$ of the antiparticles fluxes
falls monotonically with energy from a value of order 100
for kinetic energy $E \simeq 1$~GeV, 
to a value of approximately two for $E \simeq 25$~GeV, and then remains
approximately constant at the value $2.04 \pm 0.04$ in the energy
interval 30--350~GeV (up to the highest energy of the observations).
This is exactly the same behaviour of the ratio $e^+/\overline{p}$
for the production rates of antiparticles calculated using the conventional mechanism
where $e^+$ and $\overline{p}$ are created as secondary products in the
inelastic interactions of primary cosmic rays.

Many models for the CR antiparticles claim that the positrons are generated
by a new source, that is distinct and unrelated to the (conventional) mechanism that
generates antiprotons.  If this interpretation is correct, then the results described
above have to be considered as numerical accidents
of no physical significance.

The alternative is to construct a model where positrons and antiprotons
are generated by the conventional mechanism, and where the $e^+/\overline{p}$
ratio at production is preserved during propagation.
In this paper we have shown that this second possibility
is viable, and implies that the CR residence time for 
$p$, $\overline{p}$ and $e^\mp$ is of order 0.6-1.2~Myr at energy $E \simeq 1$~TeV.
This interpretation also suggests that the break in the ($e^+ + e^-$) spectrum
observed by the Atmospheric Cherenkov telescopes corresponds
to the critical energy where the residence time and the loss time for
CR $e^\mp$ are equal.

Several authors (see for example \cite{Yuksel:2008rf}) have discussed models
where the hard positron spectrum has it origin in the presence of a near source
of particles, in order to avoid the effects of energy losses.
These class of solution do not really address in a satisfactory way the puzzle that we are
discussing in this work, where the key observations is the close relation
between the positron and the antiproton fluxes. To preserve the ratio $e^+/\overline{p}$
at production for the flux, the space distribution of the sources of the two particles
must be approximately equal. A local source of secondary positrons
and antiprotons can be a viable solution but only if it accounts
for most of the fluxes for both particles.
If antiprotons can reach the Sun from large distances and positrons (because of large
energy losses) cannot, this would change the ratio at production,
unless the contribution from all distant sources is negligible for
both antiparticle types.

In the energy range considered in this work,
the CR electron spectrum is much softer than the proton one.
This difference in spectral shape is commonly understood 
as the consequence of  different propagation properties for the two particle types
that  generate different distortions on  $p$ and  $e^-$ spectra 
that are released  in interstellar space  by the CR accelerators
with approximately the same shape.
This interpretation is not possible if the Galactic residence time
of the CR particles is sufficiently short  so that 
electrons  (and  positrons) lose only a small amount of energy
inside the Galaxy. In this situation  propagation in the Galaxy  is
entirely controled by  magnetic effects  and does not depend on the particle mass.
In this  scenario, to explain the observed difference in spectral shape  between
high energy  electrons  and protons, it is necessary to conclude that the CR accelerators
release in interstellar space populations of  relativistic  $p$ and $e^-$
that have significantly different spectral shapes.

It should be stressed that the hypothesis that the CR sources
generate different  spectral shapes for protons and electrons
does  not  necessarily imply that the  acceleration  mechanism  is  not
``universal'' (i.e.  only rigidity dependent for ultrarelativistic particles),
but it has  important implications for the structure and time  evolution of
the sources, since it requires that energy losses (during or after acceleration)
must  play an important role, distorting the spectra  of the particles that
are  released in interstellar space.
As an analogy, one can  think about the Milky Way,
seen as a single accelerator
that releases in intergalactic space  populations of protons and electrons
that have very different spectral shapes.
In the standard  framework,  this difference
is attributed to the effects of the energy losses
on electrons during  their propagation  in interstellar space,
while one fundamental acceleration  mechanism creates spectra of  electrons
and protons  that have the same  energy  distributions.
Similarly one can speculate that energy losses inside individual
sources  (such as supernova  remnamnts)  can distort the spectra  of electrons
generated by the acceleration mechanism,
so that the  relativistic $e^-$ and $p$  accelerated by the source,
that reach interstellar  space
have different energy distributions.

The important conclusion is that the interpretation of the  CR positrons
as secondaries has implications of profound  importance about the properties 
of the Galactic CR  accelerators, because it requires  that these accelerators
generate spectra of  protons and electrons of different shape.
This suggests  (and in fact requires) a program of
detailed theoretical and observational studies to test the hypothesis.

In summary,  the  identification of 
the mechanisms that generate the CR positrons and antiprotons is
intimately tied to the problem of establishing the 
properties of Galactic propagation for relativistic particles,
and to the problem of the determination of the source spectra generated
by the Galactic accelerators.
Several lines of experimental studies have the potential
to contribute to  the solution for these problems:

\begin{itemize}

\item
  The simplest idea is to extend  the measurements of the positron
  and fluxes to a broader energy range.
  If the hypothesis  that the positrons have  a secondary origin
  is correct, then  the two spectra should exhibit the  same spectral
  softening  feature.  This  spectral break    has  been  already
  observed by the  Cherenkov telescopes   in the combined spectrum
  of ($e^- + e^+$)  at $E \simeq  900$~GeV (where the flux is  dominated
  by electrons).
  In the standard framework  the shape of the positron spectrum
  is determined by the properties of the new source and can have
  a variety of  different shapes.

\item 
  If the hypothesis  that the positrons have  a secondary origin
  is correct,  the space distributions of protons, electrons and positrons
  should have  approximately equal shapes.  This  is  because the
  energy losses during propagation should  be of negligible importance 
  for all three particle types, and the space distribution of the sources
  for the three  particle types are similar (but not identical).

  In the standard framework, where  the energy losses of $e^\mp$ are  important
  the space  distributions of  protons  than for   electrons,
  and the shape of  the $e^-$  energy spectra  should   be different in  different
  Galactic regions (with softer spectra  for points   away from the Galactic disk).

  Information about the density and spectral shape of CR  $p$ and
  $e^\mp$ in the Galaxy can be obtained from a study of the angular
  and energy distribution of the diffuse gamma--ray flux \cite{FermiLAT:2012aa},
  and also from the mapping of the radio emission.
  The reconstruction of the energy and space distributions of CR particles
  (protons and $e^\mp$) in the Galaxy 
  is not an easy task, and requires a sufficiently accurate
  understanding of the distributions of matter, radiation and magnetic field in the 
  Galaxy, but has great potential.

  \item 
    If the cosmic ray positrons are of secondary origin, it is  necessary
    to conclude that the CR  sources release in interstellar space
    protons and electrons spectra that have very different spectral shapes.
    This  possibility, at least in principle, can be  verified (or falsified)
    by observations.

    The prospects to obtain soon useful  measurements 
    are perhaps most  attractive  if the CR sources
    are indeed  SuperNova Remnants.  Multi--wavelength  studies
    of young  SNR's in  our Galaxy  have the potential to 
    give valuable information on the size and shape of the spectra
    of relativistic particles of different type that  the  sources
    release in interstellar space.  This is  a very difficult task
    that requires  taking into account the fact that the observations
    can study only a relatively small number of  objects each having a  different age
    and a different  environment.
\end{itemize}

\vspace {0.25 cm}
As a final remark,  we would like to note  that 
this paper, in  contrast to others, does not claim to have  established
what is the origin of  positrons  and anti--protons in cosmic rays, but consider
that this remains an open problem of  fundamental importance. 
The CR  community is in confronting two  alternative models for the
origin of  CR positrons:
\begin{itemize}
\item [(i)]
  In the most commonly accepted  framework  to describe Galactic  cosmic rays,
where the Galactic accelerators  create spectra of electrons and protons
that have  approximately the same  shape
in  a broad range of energy, it is inevitable to postulate the existence of
a new  (yet unidentified) hard source of positrons.
It is however  rather extraordinary that the new source
generates  an $e^+$  spectrum that, after distortion for propagation,
is equal (within systematic uncertainties) both in size and spectral  shape
to the spectrum of secondary  positrons that should accompany
the flux of CR anti--protons if one assumes  that all anti--protons
are  of secondary origin and that the $e^+$ energy losses are negligible.

\item[(ii)]
The striking result discussed above  naturally suggests the hypothesis
that positrons and anti--protons are both of secondary origin
(eliminating the need for a new source of uncertain nature), and that the energy
losses of  the positrons are in fact  neglible, i.e.  that the residence time
of the $e^\pm$ in the Galaxy  is sufficiently short.
This conclusion  has however important and broad implications, some of  which
go against ideas that are commonly accepted in high energy astrophysics
(for example that the source spectra of  $e^-$ and $p$  have similar shapes
or that $e^\mp$ can only propagate for  short distances in the Galaxy).
\end{itemize}
Finding the correct solution in this alternative
(positrons of secondary origin  or positrons  from a new source)
is of central importance for high energy astrophysics.


\clearpage

\appendix

\section{Analytic estimate of the $e^+/\overline{p}$ ratio at production} 
\label{sec:analytic}
If a power law spectrum of protons ($N_p(E) \simeq K_p\; E^{-\gamma_p}$)
interacts with a diluted gas target, the spectra of high energy
positrons and antiprotons generated as secondary in the interactions
have the following properties:
\begin{itemize}
\item [(a)] The $e^+$ and $\overline{p}$ spectra
 [$\dot{N}_{e^+} (E)$ and $\dot{N}_{\overline{p}} (E)$]
 in good approximation, have also power law form,
 with exponents that are approximately
 equal to each other and also 
 equal to the spectral index of the parent proton spectrum.

\item [(b)] The constant $e^+/\overline{p}$ ratio
 has a value that depends on the spectral index of the
 proton flux. For $\gamma_p \simeq 2.8 \pm 0.1$ (the exponent
 of the CR protons observed at the Earth),
 the ratio takes the value $e^+/\overline{p} \simeq 1.8 \pm 0.5$
 where the uncertainty takes also into account an estimate of 
 the uncertainties in the modeling of hadronic interactions.
\end{itemize}
This appendix outlines a simple analytic derivation
of these two results.

Point (a) is a well known ``textbook'' result, that is in fact valid
for all secondary products of hadronic inelastic collisions.
It can derived as a consequence of the approximate validity
of Feynman scaling in the forward kinematical region
for the inclusive spectra of secondaries
in high energy hadronic interactions.
%

The rate of production of secondaries of type $j$ in
$pp$ interactions can be calculated performing the integral:
 \begin{equation}
\dot{N}_{pp\to j} (E) = n_{p}^{\rm target} 
 ~\int_{E}^\infty dE_0 ~N_p (E_0) \; \beta \, c \, \sigma_{pp} (E_0)
 \; \frac{dN_{pp \to j}}{dE} \left ( E, E_0 \right ) ~.
 \label{eq:qpp-app}
 \end{equation}
 where $N_p(E_0)$ is the spectrum of interacting protons,
 $n_{p}^{\rm target}$ is the density of target protons,
 $\sigma_{pp}(E_0)$ is the inelastic $pp$ cross section,
 and $dN_{pp \to j}/dE$ is the inclusive spectrum
of secondaries of type $j$ produced in one inelastic collision.
Without loss of generality one can write the inclusive spectra 
in the form:
\begin{equation}
 \frac{dN_{pp\to j}}{dE} (E, E_0) = \frac{1}{E_0} \; 
 F_{pp\to j} \left (
 \frac{E}{E_0}, E_0 \right )~.
\end{equation}
 The approximate validity of Feynman scaling implies that 
 when the ratio $x= E/E_0$ not too small (i.e. in the forward region),
 the function $F_{pp\to j}$ depends only on $x$ (and is independent from
 $E_0$).
 For a primary proton spectrum that is a power law: 
 $N_p (E_0) = K_p \; E_0^{-\gamma_p}$ setting $\beta \simeq 1$,
 and neglecting the logarithmic energy depdendence of $\sigma_{pp}(E_0)$
 one can rewrite equation (\ref{eq:qpp-app}) as:
 \begin{equation}
 \dot{N}_{pp\to j} (E) = K_p \; n_p^{\rm target} \;
 c \; \sigma_{pp} ~Z_{pp\to j} (\gamma) ~E^{-\gamma_p}
 = K_j ~E^{-\gamma_p}
 \label{eq:qppz}
 \end{equation}
 where the ``$Z$--factor'' $Z_{pp \to j} (\gamma)$ is: 
 \begin{equation}
 Z_{pp \to j} (\gamma, E) = \int_0^1 dx~x^{\gamma -1}
 ~F_{pp \to j} \left (x\right )
 \end{equation}
 The final result is that the production rate of particle $j$ is
 described by a power law with spectral index $\gamma_j = \gamma_p$,
 as stated in point (a).

 It should be noted that this result is not exact, because it has been
 obtained making use of two approximations: the $pp$ inelastic
 cross section $\sigma_{pp}$ is constant, and the distribution $F_{pp\to j}$
 is exactly scaling.
 A numerical calculation performed without these simplifying approximations
 results in secondary spectra with small deviations from an exact power law.

 A realistic calculation must also include the fact that the parent proton
 spectrum is not an exact power law. In fact the CR proton and helium
 spectra have a spectral hardening
 (with a variation of the spectral index of order $\Delta \gamma_p \sim 0.15$) 
 at a rigidity of order 200--300~GV
 \cite{pamela-protons-helium,ams-protons,Aguilar:2016kjl}
 Secondaries of energy $E$ are injected by a broad distribution
 of primary energies that extends from $E_0$ just above $E$
 to $E_0 \approx 100~E$.
 The spectral structure of the primary particles generates
 also a similar, but more gradual hardening for the secondary spectra.

\vspace{0.2 cm}
Using the same approximations discussed above,
the ratio between the positron and anti-proton production rates
is given by the ratio of the corresponding $Z$--factors.
\begin{equation}
 \frac{\dot{N}_{pp\to e^+}}{\dot{N}_{pp\to \overline{p}}} \simeq
 \frac {Z_{pp\to e^+}(\gamma_p)}{Z_{pp\to \overline{p}}(\gamma_p)}
\end{equation}

The production of positrons in hadronic interactons
is dominated by the production and chain decay of mesons, 
and the function $F_{pp \to e^+}(x) $ can be written as a sum of 
terms that describe the production and decay of different parent mesons:
\begin{equation}
 F_{pp \to e^+} (x) = \sum_{h \in \{ \pi^+, K^+, K_L \} } 
F_{pp \to h \to e^+}(x) =
\sum_{h\in \{ \pi^+, K^+, K_L \} } 
 F_{pp \to h} (x) \otimes F_{h \to e^+}(x) ~.
\end{equation}
In this equation the symbol $\otimes$ indicate convolution and
the functions $F_{i \to e^+}(x)$ are the inclusive spectra of
positrons in the decay
of parent meson of type $i$. For ultrarelativistic
particles these functions
depend only on the ratio $E/E_i$ between the positron and the
parent meson energies.

The $Z$--factor for positron production in $pp$ interactions
can then be written as a sum of products:
 \begin{equation}
 Z_{pp \to e^+} (\gamma) = \sum_{i\in \{ \pi^+, K^+, K_L \} }
 Z_{pp \to i}(\gamma) ~Z_{i \to e^+} (\gamma)
 \end{equation}
 where the decay $Z$--factors are:
 \begin{equation}
 Z_{i \to e^+} (\gamma) = \int_0^1 dx ~x^{\gamma -1} \; F_{i\to e^+} (x) ~.
 \end{equation}
 The decay spectrum of positrons in the chain decay of $\pi^+$ can be calculated
 analytically. The calculation involves using 
 the non--trivial matrix element for 3--body muon decay, and requires to take 
 into account the fact that the muon created in the first decay is 
 in a well defined polarization state. 
 The corresponding $Z_{\pi^+ \to \mu^+ \to e^+}$ factor 
 has also an exact analytic expression \cite{Lipari:2007su}:
 \begin{equation}
 Z_{\pi^+ \to e^+} (\gamma) =
 \frac{4 \left[(\gamma (r_\pi-1)+2 r_\pi-3) r_\pi^\gamma-2 r_\pi+3 \right]}
 {\gamma^2 (\gamma+2) (\gamma+3) (r_\pi-1)^2} 
 \label{eq:zpidec}
 \end{equation}
 with $r_\pi = (m_\mu/m_\pi)^2$.
 Similarly it is possible to calculate the decay spectra and the
 corresponding $Z$ factors for the decay of kaons.

 For an exponent $\gamma = 2.8$ (a value close to the observed spectrum of 
 hadronic cosmic rays) the decay $Z$ factors for the different mesons are: 
 $Z_{\pi^+ \to e^+} \simeq 0.1219$, $Z_{K^+ \to e^+} \simeq 0.0563$,
 $Z_{K_L \to e^+} \simeq 0.0354$ (also $K^-$ can produce positrons
 via the chain decay $K^- \to \pi^+ \to e^+$, but the channel
 is strongly suppressed $Z_{K^- \to e^+} \simeq 0.0010$).
 The positrons created in pion decay have a harder spectrum,
 and therefore a larger $Z$--factor.
 The production of kaons in hadronic interaction is suppressed by
 approximately a factor of 10 with respect to the production of pions. 
 Combining this fact with the softer spectra of the positrons
 created in kaon decay (that results in additional suppression by a factor 2 or 3)
 one finds that approximately 90\% of the positron injection
 is due to the production of positive pions, with the rest due to the
 kaon contribution.

 The inclusive differential cross section for the production of 
 antiprotons and pions can be approximately fitted 
\cite{Brenner:1981kf,Badhwar:1977zf} with the simple functional form:
 \begin{equation}
 F_{pp \to h} (x) \simeq A_h \; \frac{(1-x)^{\eta_h}}{x}
 \simeq \langle x_h \rangle \; (\eta_h +1)\;
 \frac{(1-x)^{\eta_h}}{x}
 \label{eq:scal-brenner}
 \end{equation}
 that depends on the two parameters \{$A_h, \eta_h\}$,
 or equivalently $\{ \left \langle x_h \right \rangle, \eta_h \}$.
 The quantity
 $ \langle x_h\rangle = \left \langle E_h \right \rangle/E_0$ is the
 average fraction of the projectile energy $E_0$ carried by hadrons 
 of type $h$ in the final state.
 The corresponding $Z$--factors are:
 \begin{equation}
 Z_{pp\to h} (\gamma) = \langle x_h \rangle \; (n_h +1) ~
 \frac{\Gamma(\gamma-1) \; \Gamma (n_h +1)}{\Gamma(\gamma+ n_h)}~.
\label{eq:zhad}
 \end{equation}
One now all the elements to obtain a simple estimate of the
 $e^+/\overline{p}$ ratio: 
 \begin{equation}
 \frac{\dot{N}_{pp\to e^+} }{\dot{N}_{pp\to \overline{p}} } \simeq
 \frac {Z_{pp\to e^+}(\gamma_p)}{Z_{pp\to \overline{p}}(\gamma_p)}
 \simeq 
 \frac {Z_{pp\to \pi^+}(\gamma_p) 
\times Z_{\pi^+ \to e^+} (\gamma_p)}{Z_{pp\to \overline{p}}(\gamma_p)} \times 1.1
 \end{equation}
In this expression $\gamma_p$ is the spectral index of the proton flux, 
and the last factor 1.1 takes into account
a 10\% kaon contribution to the positron flux. 
Using the expression of equation (\ref{eq:zhad}) for $Z_{pp \to h}$ one obtains:
 \begin{equation}
 \frac{\dot{N}_{pp\to e^+}}{\dot{N}_{pp\to \overline{p}} } \simeq
 \frac {\langle x_{\pi^+} \rangle}{ \langle x_{\overline{p}} \rangle} 
~ \left [
 \frac{\Gamma(\gamma+\eta_{\overline{p}}) \; \Gamma (\eta_{\pi^+} +2)}
 {\Gamma(\gamma+\eta_{\pi^+}) \; \Gamma (\eta_{\overline{p}} +2)}
 \right ]
 ~ {Z_{\pi^+ \to e^+}(\gamma)} \times 1.1
\label{eq:estimate}
 \end{equation}
 In this expression one can identify four factors. The first one
 $\langle x_{\pi^+} \rangle/\langle x_{\overline{p}} \rangle$ is 
 the ratio of the energy fractions transfered to $\overline{p}$ or
 $\pi^+$ (the dominant source of positrons) by the primary particle.
 A rough estimate for this factor is of order $9.1 \pm 1.5$, where
 the central value has been estimated for 
 $\langle x_{\pi^+} \rangle \simeq 0.165$.
and $\langle x_{\overline{p}} \rangle \simeq 0.020$.

 The second factor (in square parenthesis and
 written as a combination of $\Gamma$ functions) takes into account the shapes
 of the primary proton spectrum and of the inclusive production
 cross sections for $\overline{p}$ and $\pi^+$. It
 depends on the spectral index $\gamma_p$ of the
 proton spectrum, and on the parameters
 $\eta_{\overline{p}}$ and $\eta_{e^+}$ that determine the shapes 
 of the energy spectra of antiprotons and positrons in the final state.
 This factors is larger than unity, because
 on average the secondary antiprotons are softer than charged pions
$(\eta_{\overline{p}} > \eta_{\pi^+}$)
 For $\gamma_p \simeq 2.8$, $\eta_{\pi^+} \simeq 3.5$ and $\eta_{\overline{p}} \simeq 7$,
 the factor is approximately 1.5. 
 Distortions of the secondary particle spectra (changes in the exponents 
$\Delta p \simeq \pm 1.0$)
 and changes in the spectral index of the primary flux $\Delta \gamma_p \simeq \pm 0.1 $
 can change the value of this factor by $\pm 0.4$.

 The third factor [$Z_{\pi^+ \to e^+}^{-1}(\gamma_p)$]
 is the kinematical suppression factor for $e^+$ production
 due to the fact that positrons carry only a fraction of the $\pi^+$ energy
 and is given in equation (\ref{eq:zpidec}).
 Assuming $\gamma_p \simeq 2.8$ with an uncertainty
 of order $\pm 0.1$ one obtains the estimate $0.12 \pm 0.01$.

 The last factor (1.1) in equation (\ref{eq:estimate}) takes into account a 
 kaon contribution to the positron flux of approximately 10\%. 

 Putting together the results outlined above one arrives to the estimate:
 \begin{equation}
 \frac{\dot{N}_{pp\to \overline{p}}}{\dot{N}_{pp\to e^+}} \simeq
 (9.1 \pm 1.5) \times (1.5 \pm 0.4) \times (0.12 \pm 0.01) \times (1.10 \pm 0.02)
 \simeq 1.80 \pm 0.5
\label{eq:ratio-num}
 \end{equation}
This completes the derivation of the prediction (b).

An estimate   of the  asymptotic
positron/antiproton ratio  at production: 
$e^+/\overline{p} \approx 3$   is given in  \cite{Katz:2009yd}.
This result  (given without an estimate of the systematic uncertainty) 
has  been calculated  for a  prediction of the the antiparticle
spectra  much softer than the data  (with asymptotic spectral  indices
$\gamma_{e^+} \simeq \gamma_{\overline{p}} \simeq 3.2$).  Correcting this estimate
to account for the flatter  observed spectra one obtains a ratio of order 2.50. 

Naively one could expect that the ratio positron/antiproton 
should be significantly larger than two, because the average number 
of positrons created in hadronic interactions 
is much larger than the number of antiprotons. 
However, we are not calculating the ratio of the total
numbers of $e^+$ and $\overline{p}$, but 
the ratio of the numbers of particles with the same energy $E$. 
The production rate of positrons is suppressed with respect
to antiprotons because the $e^+$ are produced in the chain decay
of parent mesons and have a much softer spectrum. 
The ratio $e^+/\overline{p}$ ratio of order two emerges as the combination
of a much larger positron multiplicity per interaction, and a softer positron
spectrum.

\clearpage

\clearpage

\begin{thebibliography}{200}

\bibitem{Adriani:2008zr} 
 O.~Adriani {\it et al.} [PAMELA Collaboration],
 Nature {\bf 458}, 607 (2009)
 [arXiv:0810.4995 [astro-ph]].

\bibitem{Adriani:2013uda} 
 O.~Adriani {\it et al.} [PAMELA Collaboration],
 Phys.\ Rev.\ Lett.\ {\bf 111}, 081102 (2013)
 [arXiv:1308.0133 [astro-ph.HE]].


\bibitem{pamela-antiprotons}
 O.~Adriani {\it et al.} [PAMELA Collaboration],
 Phys.\ Rev.\ Lett.\ {\bf 105}, 121101 (2010)
 [arXiv:1007.0821 [astro-ph.HE]].


\bibitem{FermiLAT:2011ab} 
 M.~Ackermann {\it et al.} [Fermi-LAT Collaboration],
 Phys.\ Rev.\ Lett.\ {\bf 108}, 011103 (2012)
 [arXiv:1109.0521 [astro-ph.HE]].

\bibitem{Aguilar:2014mma} 
 M.~Aguilar {\it et al.} [AMS Collaboration],
 Phys.\ Rev.\ Lett.\ {\bf 113}, 121102 (2014).
 
\bibitem{Aguilar:2016kjl} 
  M.~Aguilar {\it et al.} [AMS Collaboration],
  Phys.\ Rev.\ Lett.\  {\bf 117}, no. 9, 091103 (2016).
  doi:10.1103/PhysRevLett.117.091103

\bibitem{Adriani:2011xv} 
 O.~Adriani {\it et al.} [PAMELA Collaboration],
 Phys.\ Rev.\ Lett.\ {\bf 106}, 201101 (2011)
 [arXiv:1103.2880 [astro-ph.HE]].

 \bibitem{pamela-protons-helium}
 O.~Adriani {\it et al.} [PAMELA Collaboration],
 Science {\bf 332}, 69 (2011)
 [arXiv:1103.4055 [astro-ph.HE]].

 
\bibitem{ams-protons} 
 M.~Aguilar {\it et al.} [AMS Collaboration],
 Phys.\ Rev.\ Lett.\ {\bf 114}, 171103 (2015).


\bibitem{Yoon:2011aa} 
 Y.~S.~Yoon {\it et al.},
 Astrophys.\ J.\ {\bf 728}, 122 (2011)
 [arXiv:1102.2575 [astro-ph.HE]].

 
\bibitem{Abdo:2009zk} 
 A.~A.~Abdo {\it et al.} [Fermi-LAT Collaboration],
 Phys.\ Rev.\ Lett.\ {\bf 102}, 181101 (2009)
 [arXiv:0905.0025 [astro-ph.HE]].

 
\bibitem{Aharonian:2008aa} 
 F.~Aharonian {\it et al.} [HESS Collaboration],
 Phys.\ Rev.\ Lett.\ {\bf 101}, 261104 (2008)
 [arXiv:0811.3894 [astro-ph]].

\bibitem{Aharonian:2009ah} 
 F.~Aharonian {\it et al.} [HESS Collaboration],
 Astron.\ Astrophys.\ {\bf 508}, 561 (2009)
 [arXiv:0905.0105 [astro-ph.HE]].


\bibitem{cream-discrepant-hardening}
 H.~S.~Ahn {\it et al.},
 Astrophys.\ J.\ {\bf 714}, L89-L93 (2010).
 [arXiv:1004.1123 [astro-ph.HE]].


\bibitem{BorlaTridon:2011dk} 
 D.~Borla Tridon {\it et al.} [MAGIC Collaboration],
 arXiv:1110.4008 [astro-ph.HE].

\bibitem{Staszak:2015kza} 
 D.~Staszak [VERITAS Collaboration],
 arXiv:1508.06597 [astro-ph.HE].

\bibitem{Lipari:2007su} 
 P.~Lipari, M.~Lusignoli and D.~Meloni,
 Phys.\ Rev.\ D {\bf 75}, 123005 (2007)
 [arXiv:0704.0718 [astro-ph]].

\bibitem{Aguilar:2014fea} 
 M.~Aguilar {\it et al.} [AMS Collaboration],
 Phys.\ Rev.\ Lett.\ {\bf 113}, 221102 (2014).
 doi:10.1103/PhysRevLett.113.221102
 
\bibitem{Aguilar:2015ctt} 
 M.~Aguilar {\it et al.} [AMS Collaboration],
 Phys.\ Rev.\ Lett.\ {\bf 115}, no. 21, 211101 (2015).
 doi:10.1103/PhysRevLett.115.211101

\bibitem{Gaisser:2013bla} 
 T.~K.~Gaisser, T.~Stanev and S.~Tilav, \\
 arXiv:1303.3565 [astro-ph.HE].

\bibitem{Engelmann:1990zz} 
 J.~J.~Engelmann, P.~Ferrando, A.~Soutoul, P.~Goret and E.~Juliusson,
 Astron.\ Astrophys.\ {\bf 233}, 96 (1990).

\bibitem{Gleeson:1968zza} 
 L.~J.~Gleeson and W.~I.~Axford,
 Astrophys.\ J.\ {\bf 154}, 1011 (1968).

\bibitem{Lipari:2014gfa} 
 P.~Lipari,
 arXiv:1408.0431 [astro-ph.HE].


 
\bibitem{Tan:1982nc} 
 L.~C.~Tan and L.~K.~Ng,
 Phys.\ Rev.\ D {\bf 26}, 1179 (1982).


\bibitem{Badhwar:1977zf} 
 G.~D.~Badhwar, S.~A.~Stephens and R.~L.~Golden,
 Phys.\ Rev.\ D {\bf 15}, 820 (1977).

\bibitem{Anticic:2010yg} 
 T.~Anticic {\it et al.} [NA49 Collaboration],
 Eur.\ Phys.\ J.\ C {\bf 68}, 1 (2010)
 [arXiv:1004.1889 [hep-ex]].

\bibitem{Sjostrand:2006za} 
 T.~Sjostrand, S.~Mrenna and P.~Z.~Skands,
 JHEP {\bf 0605}, 026 (2006)
 [hep-ph/0603175].
 

\bibitem{Tilley:2004zz} 
 D.~R.~Tilley, J.~H.~Kelley, J.~L.~Godwin, D.~J.~Millener, J.~E.~Purcell, C.~G.~Sheu and H.~R.~Weller,
 Nucl.\ Phys.\ A {\bf 745}, 155 (2004).


\bibitem{beryllium1} 
M.~Garcia-Munoz, G.M.~Mason \& J.A.~Simpson 
Astrophys. \ J. {\bf 217}, 859 (1977).

\bibitem{beryllium2} 
S.P.~Ahlen {\it et al.}
Astrophys. \ J. \ {\bf 534}, 757 (2000).

\bibitem{beryllium3} 
N.E.~Yanasak {\it et al.}
Astrophys.\ J.\ {\bf 563}, 768 (2001).

\bibitem{Hillas:2005cs} 
 A.~M.~Hillas,
 J.\ Phys.\ G {\bf 31}, R95 (2005).
 doi:10.1088/0954-3899/31/5/R02


\bibitem{Mertsch:2014cua} 
 P.~Mertsch and S.~Funk,
 Phys.\ Rev.\ Lett.\ {\bf 114}, no. 2, 021101 (2015)
 doi:10.1103/PhysRevLett.114.021101
 [arXiv:1408.3630 [astro-ph.HE]].

\bibitem{lipari-anisotropy} 
Paolo Lipari, in preparation.
 
\bibitem{deNolfo:2006qj} 
 G.~A.~de Nolfo {\it et al.},
 Adv.\ Space Res.\ {\bf 38}, 1558 (2006)
 [astro-ph/0611301].

\bibitem{Ahn:2008my} 
 H.~S.~Ahn {\it et al.},
 Astropart.\ Phys.\ {\bf 30}, 133 (2008)
 [arXiv:0808.1718 [astro-ph]].


\bibitem{Obermeier:2012vg} 
 A.~Obermeier, P.~Boyle, J.~Horandel and D.~Muller,
 Astrophys.\ J.\ {\bf 752}, 69 (2012)
 [arXiv:1204.6188 [astro-ph.HE]].

\bibitem{Adriani:2014xoa} 
 O.~Adriani {\it et al.},
 Astrophys.\ J.\ {\bf 791}, no. 2, 93 (2014)
 [arXiv:1407.1657 [astro-ph.HE]].

\bibitem{Cowsik:2010zz} 
 R.~Cowsik and B.~Burch,
 Phys.\ Rev.\ D {\bf 82}, 023009 (2010).

\bibitem{Cowsik:2013woa} 
 R.~Cowsik, B.~Burch and T.~Madziwa-Nussinov,
 Astrophys.\ J.\ {\bf 786}, 124 (2014)
 [arXiv:1305.1242 [astro-ph.HE]].

\bibitem{Cowsik:2015yra} 
 R.~Cowsik and T.~Madziwa-Nussinov,
 arXiv:1505.00305 [astro-ph.HE].


\bibitem{ams-days-lithium}
L.~Derome, Talk at ``AMS Days at CERN'' 
available at {\tt http://indico.cern.ch/event/381134/} (2015)


\bibitem{Aguilar:2016vqr} 
  M.~Aguilar {\it et al.} [AMS Collaboration],
  Phys.\ Rev.\ Lett.\  {\bf 117}, no. 23, 231102 (2016).
  doi:10.1103/PhysRevLett.117.231102


\bibitem{Tomassetti:2015nha} 
  N.~Tomassetti,
  Phys.\ Rev.\ C {\bf 92}, no. 4, 045808 (2015)
  doi:10.1103/PhysRevC.92.045808
  [arXiv:1509.05776 [astro-ph.HE]].


\bibitem{Katz:2009yd} 
 B.~Katz, K.~Blum and E.~Waxman,
 Mon.\ Not.\ Roy.\ Astron.\ Soc.\ {\bf 405}, 1458 (2010)
 doi:10.1111/j.1365-2966.2010.16568.x
 [arXiv:0907.1686 [astro-ph.HE]].

\bibitem{Blum:2013zsa} 
 K.~Blum, B.~Katz and E.~Waxman,
 Phys.\ Rev.\ Lett.\ {\bf 111}, no. 21, 211101 (2013)
 doi:10.1103/PhysRevLett.111.211101
 [arXiv:1305.1324 [astro-ph.HE]].


\bibitem{Ahlen:2014ica} 
  M.~Kruskal, S.~P.~Ahlen and G.~Tarlé,
  Astrophys.\ J.\  {\bf 818}, no. 1, 70 (2016)
  doi:10.3847/0004-637X/818/1/70
  [arXiv:1410.7239 [astro-ph.HE]].

  
\bibitem{Donato:2008jk} 
  F.~Donato, D.~Maurin, P.~Brun, T.~Delahaye and P.~Salati,
  Phys.\ Rev.\ Lett.\  {\bf 102}, 071301 (2009)
  [arXiv:0810.5292 [astro-ph]].

\bibitem{Trotta:2010mx} 
  R.~Trotta, G.~Johannesson, I.~V.~Moskalenko, T.~A.~Porter, R.~R.~de Austri and A.~W.~Strong,
  Astrophys.\ J.\  {\bf 729}, 106 (2011)
  [arXiv:1011.0037 [astro-ph.HE]].


    
\bibitem{Giesen:2015ufa} 
  G.~Giesen, M.~Boudaud, Y.~Génolini, V.~Poulin, M.~Cirelli, P.~Salati and P.~D.~Serpico,
  JCAP {\bf 1509}, no. 09, 023 (2015)
  doi:10.1088/1475-7516/2015/09/023, 10.1088/1475-7516/2015/9/023
  [arXiv:1504.04276 [astro-ph.HE]].

\bibitem{Evoli:2015vaa} 
  C.~Evoli, D.~Gaggero and D.~Grasso,
  JCAP {\bf 1512}, no. 12, 039 (2015)
  doi:10.1088/1475-7516/2015/12/039
  [arXiv:1504.05175 [astro-ph.HE]].

\bibitem{Yuksel:2008rf} 
 H.~Yuksel, M.~D.~Kistler and T.~Stanev,
 Phys.\ Rev.\ Lett.\ {\bf 103}, 051101 (2009)
 doi:10.1103/PhysRevLett.103.051101
 [arXiv:0810.2784 [astro-ph]].
 
\bibitem{FermiLAT:2012aa} 
 [Fermi-LAT Collaboration],
 ``Fermi-LAT Observations of the Diffuse Gamma-Ray Emission: Implications for Cosmic Rays and the Interstellar Medium,''
 Astrophys.\ J.\ {\bf 750}, 3 (2012)
 [arXiv:1202.4039 [astro-ph.HE]].


\bibitem{Brenner:1981kf} 
 A.~E.~Brenner {\it et al.},
 Phys.\ Rev.\ D {\bf 26}, 1497 (1982).

\end{thebibliography}
\end{document}